\documentclass[aps]{revtex4}
\usepackage{eurosym}
\usepackage{amsfonts}
\usepackage{amsmath}
\usepackage{amssymb,epsf}
\usepackage{color}
\usepackage{xcolor}
\usepackage{graphicx}
\usepackage{epstopdf}
\usepackage{float}
\usepackage{caption}
\usepackage{subfig}
\usepackage{hyperref}
\usepackage{mathtools}
\usepackage{hyperref}
\usepackage{orcidlink}

\begin{document}

\title{Quasinormal modes and emission rate of ModMax (A)dS black holes}
\author{
Behzad Eslam Panah$^{1,2,3 \orcidlink{0000-0002-1447-3760}}$ \footnote{email address: \href{mailto:eslampanah@umz.ac.ir}{eslampanah@umz.ac.ir} },
Angel Rincon$^{4 \orcidlink{0000-0001-8069-9162}}$
\footnote{email address: \href{mailto:angel.rincon@ua.es}{angel.rincon@ua.es}}, 
Narges Heidari$^{1}$ \footnote{email address: \href{mailto:heidari.n@gmail.com}{heidari.n@gmail.com}}}
\affiliation{$^{1}$ Department of Theoretical Physics, Faculty of Science, University of
Mazandaran, P. O. Box 47416-95447, Babolsar, Iran}
\affiliation{$^{2}$ ICRANet-Mazandaran, University of Mazandaran, P. O. Box 47416-95447
Babolsar, Iran}
\affiliation{$^{3}$ ICRANet, Piazza della Repubblica 10, I-65122 Pescara, Italy}
\affiliation{$^{4}$ Departamento de F{\'i}sica Aplicada, Universidad de Alicante, Campus
de San Vicente del Raspeig, E-03690 Alicante, Spain}

\begin{abstract}
By considering a new model of nonlinear electrodynamics, known as the
modified Maxwell (ModMax), and taking into account the topological and the
cosmological constants in Einstein's gravity, we extract black hole
solutions called Topological ModMax (A)dS black holes. The next step is to
study the thermodynamic properties, quasinormal modes, and emission rates of
these black holes in order to examine the impact of ModMax's parameter and
the cosmological constant on these systems. To achieve this, we obtain the
quasinormal spectra for massless scalar, electromagnetic, and Dirac
perturbations. Additionally, we calculate null geodesics and determine the
radius of the critical orbit. We then apply this information to derive the
angular velocity and the Lyapunov exponent, which represent the real and
imaginary terms of the quasinormal modes in the eikonal limit, respectively.
Furthermore, we investigate the energy emission rate based on the discussion
of null geodesics and the shadow radius.
\end{abstract}

\maketitle

\section{\textbf{Introduction}}

The nonlinear electrodynamics (NED) theories were introduced to address some
of the problems in Maxwell's theory, such as the singularity that occurs at
the position of a point-like charge. This singularity is one of the main
issues in Maxwell's theory. One of the most well-known NED theories is the
Born-Infeld (BI) theory \cite{BI}. The BI theory arises as the effective
action in an open super-string theory and D-brains. It has a non-singular
self-energy of the point-like charge (see Refs. \cite{BI1, BI2} for more
details). However, the BI-NED theory is less symmetric than Maxwell's
theory. Although this NED theory possesses a fundamental SO(2) duality \cite%
{BI3}, it features a dimensionful parameter and, therefore, is no longer
conformally invariant. On the other hand, Maxwell's theory is the only
theory with two fundamental symmetries: electromagnetic duality and
conformal invariance. Recently, a new model of NED was introduced in Refs. 
\cite{ModMaxI, ModMaxII}, which includes the same symmetries as Maxwell's
theory. This NED theory is known as the Modification of Maxwell (ModMax)
theory. The ModMax theory is characterized by a dimensionless parameter $%
\gamma$ (ModMax parameter) and reduces to Maxwell's theory when $\gamma=0$
(see Refs. \cite%
{ModMaxBH,MM1,MM2,MM3,MM4,MM5,MM6,MM7,MM8,MM9,MM10,MM11,MM12,MM13,MM14,MM15,MM16,MM17,MM18,MM19,MM20,MM21,MM22,MM23,MM24,MM25,MM26}
for more details on the ModMax theory).

Bekenstein and Hawking discovered an analogy between the geometric
properties of black holes and thermodynamic variables that provides a
profound understanding of the relationship between the physical properties
of gravity and classical thermodynamics \cite{BekHaw}. The thermodynamic
properties of black holes, particularly phase transition and thermal
stability, are of great importance as they offer insights into the
underlying structure of spacetime geometry. Phase transition also plays a
significant role in various fields such as elementary particles \cite%
{Kleinert}, usual thermodynamics \cite{Callen}, condensed matter \cite{Greer}%
, black holes \cite{Zou}, and cosmology \cite{Layzer}. Another intriguing
aspect of black hole thermodynamics is the phase transition in anti-de
Sitter (AdS) spacetime, which is motivated by AdS/CFT duality \cite{Witten}%
), and provides valuable insights into the quantum nature of gravity. The
study of phase transition in AdS black holes within an extended phase space
has gained considerable attention in recent years. In this perspective, the
cosmological constant can be considered as a variable, identified as
thermodynamic pressure \cite{LambdaPI}, while the black hole's mass is
interpreted as enthalpy \cite{LambdaPII}.

Isolated black holes are ideal objects and are considered starting points
for studying a more realistic situation. Black holes are never isolated, so
they interact with their surroundings and perturb the background. After such
an interaction, the black hole responds by emitting gravitational waves,
which are parameterized by (quasinormal) modes with characteristic
frequencies. Such modes have a non-vanishing imaginary part, and they
contain all the information about how the black holes relax after the
perturbation. It should be emphasized that the quasinormal frequencies are
affected by i) the geometry itself, and ii) the type of perturbation
"applied" to the background (scalar, vector, tensor, or fermionic), and they
are independent of the initial conditions.

During the ring-down phase of a black hole merger, a unique perturbed object
is created. This is when the geometry of spacetime undergoes damped
oscillations due to the emission of gravitational waves. In this context,
quasi-normal modes (QNMs) play a significant role. To understand the
underlying physics of gravitational waves, black hole perturbation theory is
essential. Several references such as \cite%
{Regge:1957td,Zerilli:1970se,Zerilli:1970wzz,Zerilli:1974ai,Moncrief:1975sb,Teukolsky:1972my}
provide valuable insights in this field,
while more recent work, in which a wide variety of methods and a wide range of backgrounds have been investigated, can be found in \cite{Alfaro:2024tdr,Becar:2022wcj,Gonzalez:2022ote,Fernando:2022wlm,Breton:2017hwe,Fernando:2016ftj,Rincon:2024won,Gogoi:2023fow,Baruah:2023rhd,Gogoi:2023kjt,Ovgun:2023ego,Pantig:2022gih,Fontana:2024odl,DalBoscoFontana:2023syy,Panotopoulos:2017hns,Rincon:2018sgd,Destounis:2018utr,Rincon:2018ktz,Panotopoulos:2019qjk,Panotopoulos:2019gtn,Rincon:2020iwy,Rincon:2020cos,Panotopoulos:2020mii,Balart:2023swp,Rincon:2023hvd,Balart:2023odm,Konoplya:2003ii,Konoplya:2019hlu,Cardoso:2003cj} and references therein.
Gravitational wave astronomy
provides a powerful tool for testing gravity under extreme conditions. While
there is a vast amount of literature available, some excellent reviews on
the subject can be found at \cite%
{Kokkotas:1999bd,Berti:2009kk,Konoplya:2011qq}. Additionally,
Chandrasekhar's seminal monograph \cite{Chandrasekhar:1985kt} delves into
non-trivial details, covering the canonical aspects of black hole physics.

The study of QNMs is relevant not only for their constraints on the
parameters of the black hole but also for their connection with area
quantization. Therefore, it is now more important than ever to investigate
the QN spectrum of black holes, which is precisely one of the main goals of
this manuscript.

On the other hand, a method of analytically evaluating the QNMs in the
geometric-optics (eikonal) limit has been introduced in Ref. \cite{mashhoon}%
. Furthermore, Ref. \cite{cardoso} demonstrates that in this eikonal limit,
QNMs are associated with the properties of a null particle situated on the
unstable circular geodesic of the spacetime. This relationship has been
confirmed in the majority of static, spherically symmetric, asymptotically
flat spacetimes. The real part of the QNM is interpreted as the angular
velocity at the unstable null geodesic, while the imaginary part is
connected to the Lyapunov exponent, corresponding to the instability time
scale of the orbit \cite{kimet1,lopez1,giri,mondal,abbas}.

After a brief and concise introduction, we proceed to introduce and
summarize the relevant equations within the framework of ModMax theory and
topological black hole solutions. Moving on to section III, we delve into
the investigation of black hole thermodynamics, including the calculation of
thermal stability. Section IV is dedicated to the computation of the
corresponding QNMs for i) scalar, ii) electromagnetic, and iii) Dirac
perturbations. The analysis of QNMs in eikonal limits is presented in
section V. Subsequently, in section VI, we delve into the discussion of the
energy emission rate for various parameters of the ModMax black hole.
Finally, we conclude and discuss our findings in the last section.

\section{ModMax Theory and Topological Black Hole Solutions}

The action describing the coupling of Einstein's gravity with the ModMax
electrodynamics and the cosmological constant is expressed as 
\begin{equation}
\mathcal{I}=\frac{1}{16\pi }\int_{\partial \mathcal{M}}d^{4}x\sqrt{-g}\left[
R-2\Lambda -4\mathcal{L}\right] ,  \label{Action}
\end{equation}%
where $R$ and $\Lambda $ are, respectively, the Ricci scalar and the
cosmological constant. The determinant of the metric tensor $g_{\mu \nu }$
is defined as $g=\text{det}(g_{\mu \nu })$. In the above action, $\mathcal{L}
$ refers to ModMax's Lagrangian and is given \cite{ModMaxI,ModMaxII} 
\begin{equation}
\mathcal{L}=\mathcal{S}\cosh \gamma -\sqrt{\mathcal{S}^{2}+\mathcal{P}^{2}}%
\sinh \gamma ,  \label{ModMaxL}
\end{equation}%
where $\gamma $ is referred to as ModMax's parameter, which is a
dimensionless quantity. Additionally, $\mathcal{S}$ and $\mathcal{P}$
represent a true scalar and a pseudoscalar, respectively, in the following
expressions 
\begin{eqnarray}
\mathcal{S} &=&\frac{\mathcal{F}}{4}, \\
&&  \notag \\
\mathcal{P} &=&\frac{\widetilde{\mathcal{F}}}{4},
\end{eqnarray}%
in the equations above, the term $\mathcal{F}=F_{\mu \nu }F^{\mu \nu }$ is
referred to as the Maxwell invariant. Moreover, $F_{\mu \nu }$ represents
the electromagnetic tensor field and is defined as $F_{\mu \nu }=\partial
_{\mu }A_{\nu }-\partial _{\nu }A_{\mu }$, with $A_{\mu }$ being the gauge
potential. Additionally, $\widetilde{\mathcal{F}}=$ $F_{\mu \nu }\widetilde{F%
}^{\mu \nu }$, where $\widetilde{F}^{\mu \nu }=\frac{1}{2}\epsilon _{\mu \nu
}^{~~~\rho \lambda }F_{\rho \lambda }$. It is worth noting that the
Lagrangian of ModMax (Eq. (\ref{ModMaxL})) reduces to the Maxwell theory
(i.e., $\mathcal{L}=\frac{\mathcal{F}}{4}$) when $\gamma =0$.

In this work, we are interested in studying electrically charged black holes
within the framework of Einstein's theory, taking into account the presence
of the cosmological constant. Therefore, we can set $\mathcal{P}=0$ in the
Lagrangian of ModMax (Eq.(\ref{ModMaxL})).

The generalized Einstein-ModMax's equations in the presence of the
cosmological constant write in the following form \cite{ModMaxBH} 
\begin{eqnarray}
G_{\mu \nu }+\Lambda g_{\mu \nu } &=&8\pi \mathrm{T}_{\mu \nu },  \label{eq1}
\\
&&  \notag \\
\partial _{\mu }\left( \sqrt{-g}\widetilde{E}^{\mu \nu }\right) &=&0,
\label{eq2}
\end{eqnarray}%
where $\mathrm{T}_{\mu \nu }$ is the energy-momentum tensor which is given
by 
\begin{equation}
8\pi \mathrm{T}^{\mu \nu }=2\left( F^{\mu \sigma }F_{~~\sigma }^{\nu
}e^{-\gamma }\right) -2e^{-\gamma }\mathcal{S}g^{\mu \nu },  \label{eq3}
\end{equation}%
and $\widetilde{E}_{\mu \nu }$ is defined as 
\begin{equation}
\widetilde{E}_{\mu \nu }=\frac{\partial \mathcal{L}}{\partial F^{\mu \nu }}%
=2\left( \mathcal{L}_{\mathcal{S}}F_{\mu \nu }\right) ,  \label{eq4}
\end{equation}%
where $\mathcal{L}_{\mathcal{S}}=\frac{\partial \mathcal{L}}{\partial 
\mathcal{S}}$.

For charged case, the ModMax field equation (Eq. (\ref{eq2})), turns to 
\begin{equation}
\partial _{\mu }\left( \sqrt{-g}e^{-\gamma }F^{\mu \nu }\right) =0.
\label{Maxwell Equation}
\end{equation}

Notably, by setting $\gamma \neq 0$ a birefringence phenomenon occurs \cite%
{ModMaxI}; apart from the lightlike polarization mode there exists another
mode which is subluminal for $\gamma >0$ and superluminal for $\gamma <0$,
hinting on a physical restriction $\gamma \geq 0$.

Here, we consider a four-dimensional static metric in the following form 
\begin{equation}
ds^{2}=-f\left( r\right) dt^{2}+\frac{dr^{2}}{f(r)}+r^{2}d\Omega _{k}^{2},
\label{Metric}
\end{equation}%
where $f(r)$ is the metric function in which we must find it. In addition, $%
d\Omega _{k}^{2}$ is given by 
\begin{equation}
d\Omega _{k}^{2}=\left\{ 
\begin{array}{ccc}
d\theta ^{2}+\sin ^{2}\theta d\varphi ^{2} &  & k=1 \\ 
d\theta ^{2}+d\varphi ^{2} &  & k=0 \\ 
d\theta ^{2}+\sinh ^{2}\theta d\varphi ^{2} &  & k=-1%
\end{array}%
\right. .
\end{equation}

To have a radial electric field, we use the following gauge potential 
\begin{equation}
A_{\mu }=h(r)\delta _{\mu }^{t},  \label{gauge
potential}
\end{equation}%
and by applying the metric (\ref{Metric}) and the MaxMax field equation (\ref%
{Maxwell Equation}), we can get 
\begin{equation}
2h^{\prime }(r)+rh^{\prime \prime }(r)=0,  \label{heq}
\end{equation}%
where the prime and double prime are, respectively, the first and second
derivatives with respect to $r$. We can solve the above equation which leads
to 
\begin{equation}
h(r)=-\frac{q}{r},  \label{h(r)}
\end{equation}%
where $q$ is an integration constant related to the electric charge.

In order to get the metric function, $f(r)$, we apply the equations (\ref%
{eq1}), (\ref{eq3}), (\ref{Metric}), and (\ref{h(r)}). After some
calculation, we can obtain the following differential equations 
\begin{eqnarray}
&&eq_{tt}=eq_{rr}=rf^{\prime }(r)+f\left( r\right) +\Lambda r^{2}-k+\frac{%
q^{2}}{r^{2}}e^{-\gamma }=0,  \label{eqENMax1} \\
&&  \notag \\
&&eq_{\theta \theta }=eq_{\varphi \varphi }=f^{\prime \prime }(r)+\frac{2}{r}%
f^{\prime }(r)+2\Lambda -\frac{2q^{2}}{r^{4}}e^{-\gamma }=0,
\label{eqENMax2}
\end{eqnarray}%
which $eq_{tt}$, $eq_{rr}$, $eq_{\theta \theta }$, and $eq_{\varphi \varphi
} $ are representative $tt$, $rr$, $\theta \theta $, and $\varphi \varphi $
components of Eq. (\ref{eq1}), respectively. Using the differential
equations (\ref{eqENMax1}) and (\ref{eqENMax2}), we extract the metric
function in the following form 
\begin{equation}
f(r)=k-\frac{m}{r}-\frac{\Lambda r^{2}}{3}+\frac{q^{2}e^{-\gamma }}{r^{2}},
\label{f(r)}
\end{equation}%
where $m$ is an integration constant that is related to the geometrical mass
of the black hole. It is notable that the obtained metric function (\ref%
{f(r)}) satisfies all the components of the field equation (\ref{eq1}),
simultaneously. Moreover, the metric function Eq. (\ref{f(r)}) turns to
topological Reissner-Nordstrom (A)dS black hole when $\gamma =0$, i.e. 
\begin{equation}
f(r)=k-\frac{m}{r}-\frac{\Lambda r^{2}}{3}+\frac{q^{2}}{r^{2}}.
\end{equation}

To find the singularity of the obtained solutions (\ref{f(r)}), we calculate
the Kretschmann scalar. For this aim, by considering the metric (\ref{Metric}%
) and the metric function (\ref{f(r)}), we can obtain the Kretschmann scalar
in the following form 
\begin{equation}
R_{\alpha \beta \gamma \delta }R^{\alpha \beta \gamma \delta }=\frac{%
8\Lambda ^{2}}{3}+\frac{12m^{2}}{r^{6}}-\frac{48mq^{2}e^{-\gamma }}{r^{7}}+%
\frac{56q^{4}e^{-2\gamma }}{r^{8}},
\end{equation}%
which indicates that the Kretschmann scalar diverges at $r=0$ (i.e., $%
\lim_{r\longrightarrow 0}R_{\alpha \beta \gamma \delta }R^{\alpha \beta
\gamma \delta }\longrightarrow \infty $). Therefore, there is a curvature
singularity located at $r=0$. Also, it is finite for $r\neq 0$. The
asymptotic behavior of spacetime is determined by $\lim_{r\longrightarrow
\infty }R_{\alpha \beta \gamma \delta }R^{\alpha \beta \gamma \delta
}\longrightarrow \frac{8\Lambda ^{2}}{3}$, and $\lim_{r\longrightarrow
\infty }f\left( r\right) \longrightarrow -\frac{\Lambda r^{2}}{3}$, where
shows the spacetime will be asymptotically de sitter (dS) or anti de Sitter
(AdS), for $\Lambda >0$ (or $\Lambda <0$).

\begin{figure}[tbph]
\centering
\includegraphics[width=0.35\linewidth]{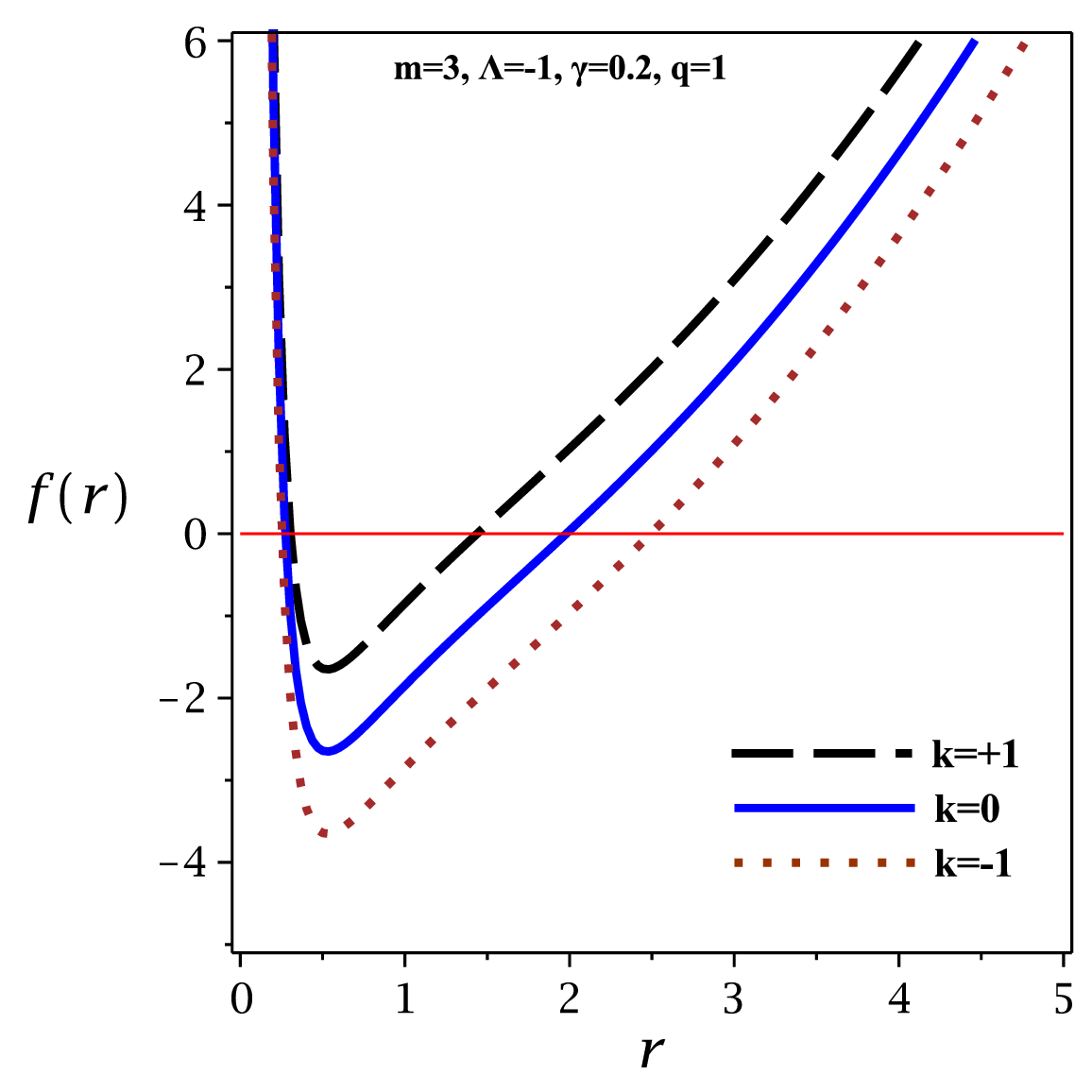} \includegraphics[width=0.35%
\linewidth]{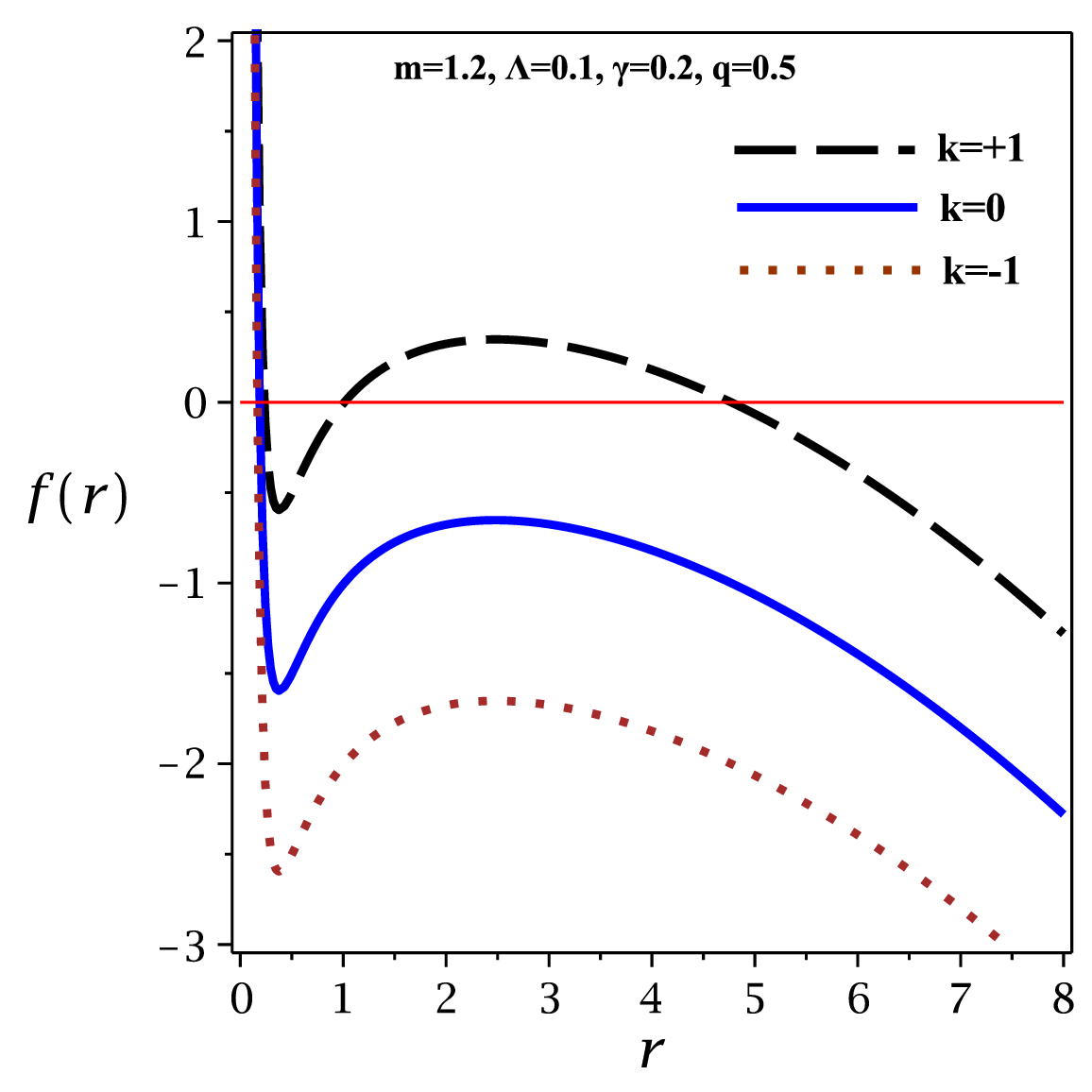} \newline
\caption{The metric function $f(r)$ versus $r$ for different values of the
topological constant ($k$). Left panel for AdS case ($\Lambda<0$). Right
panel for dS case ($\Lambda>0$).}
\label{Fig1}
\end{figure}

\begin{figure}[tbph]
\centering
\includegraphics[width=0.35\linewidth]{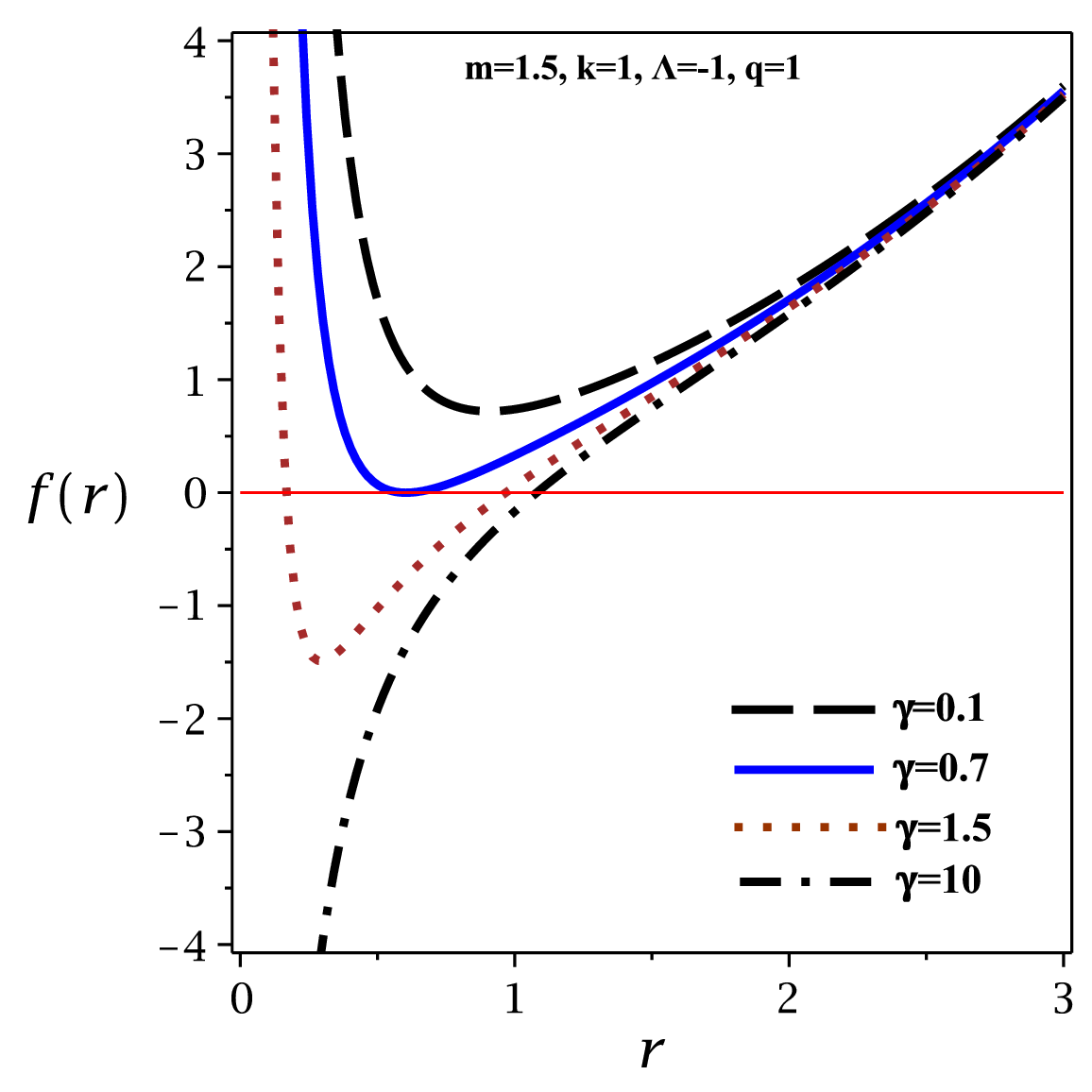} \includegraphics[width=0.35%
\linewidth]{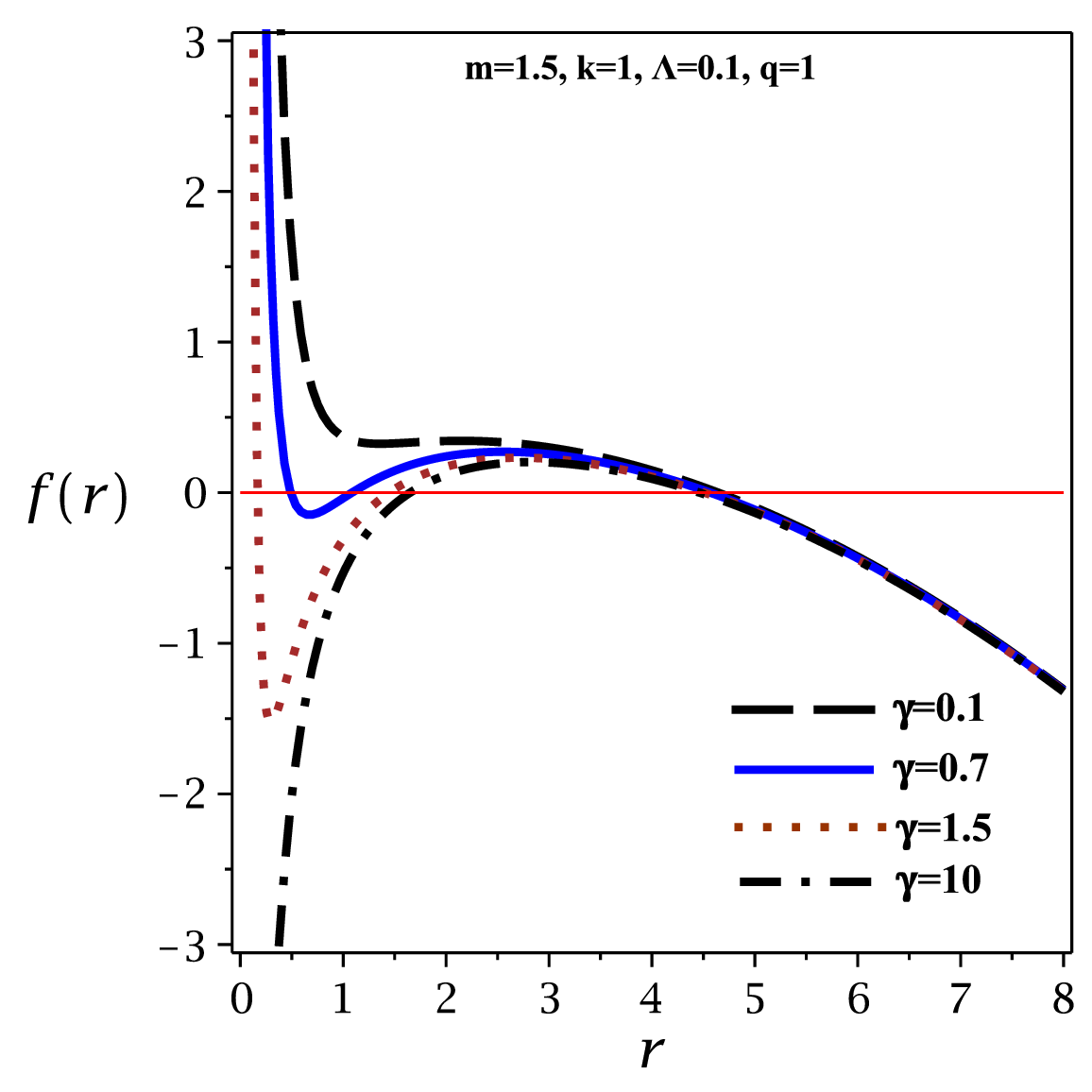} \newline
.
\caption{The metric function $f(r)$ versus $r$ for $k=+1$, and different
values of the ModMax's parameter. Left panel for AdS case ($\Lambda<0$).
Right panel for dS case ($\Lambda>0$)}
\label{Fig2}
\end{figure}

\begin{figure}[tbph]
\centering
\includegraphics[width=0.35\linewidth]{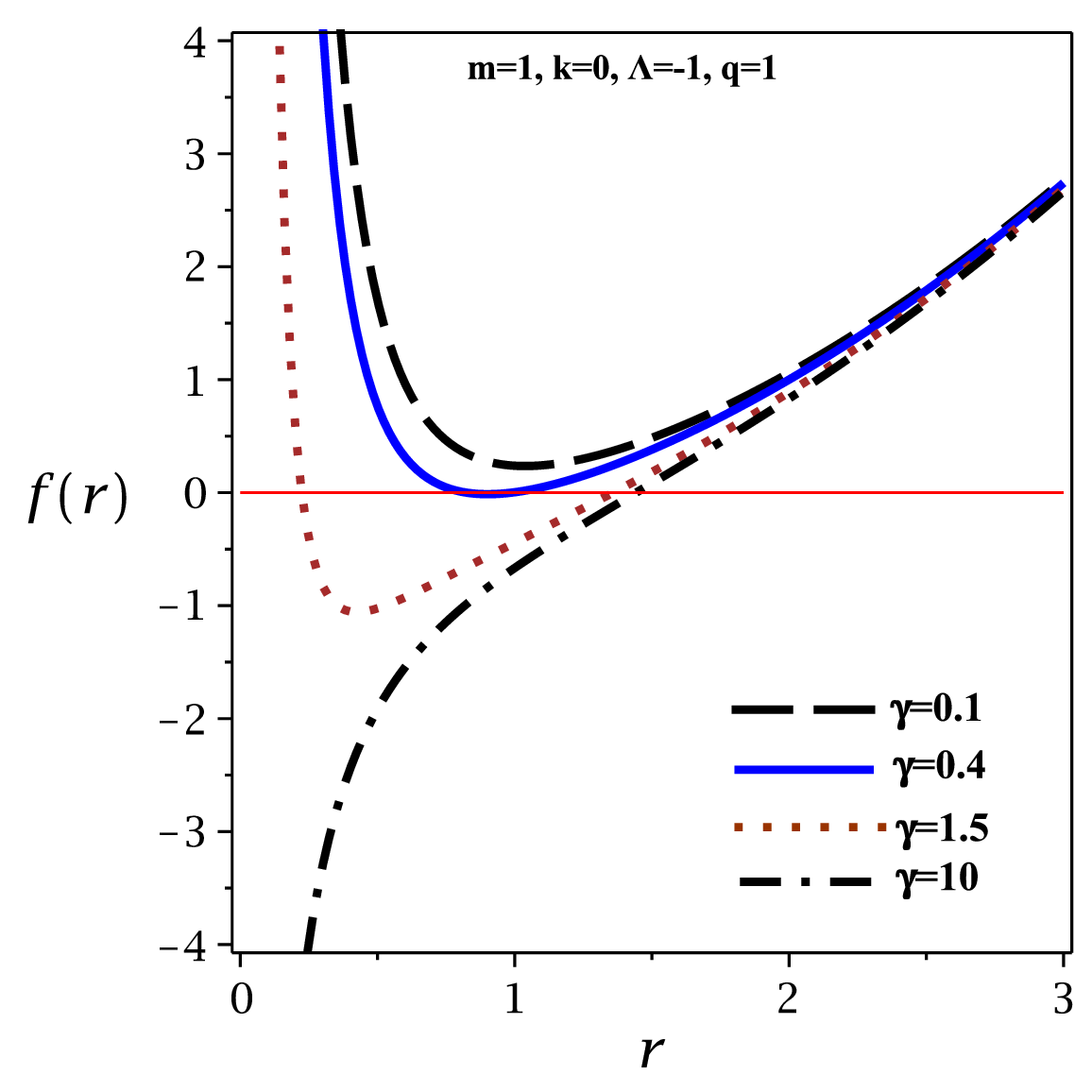} \includegraphics[width=0.35%
\linewidth]{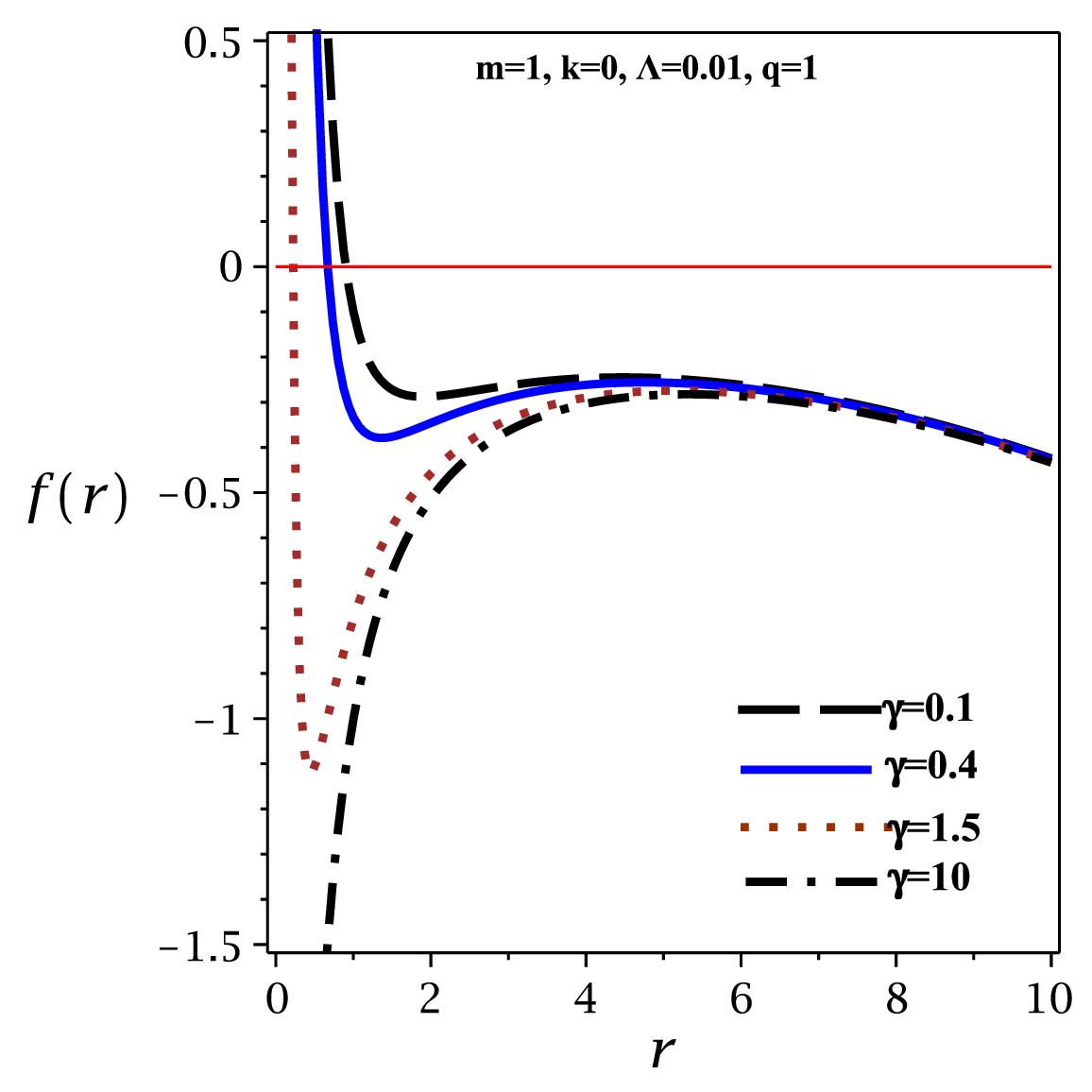} \newline
\caption{The metric function $f(r)$ versus $r$ for $k=0$, and different
values of the ModMax's parameter. Left panel for AdS case ($\Lambda<0$).
Right panel for dS case ($\Lambda>0$).}
\label{Fig3}
\end{figure}

\begin{figure}[tbph]
\centering
\includegraphics[width=0.35\linewidth]{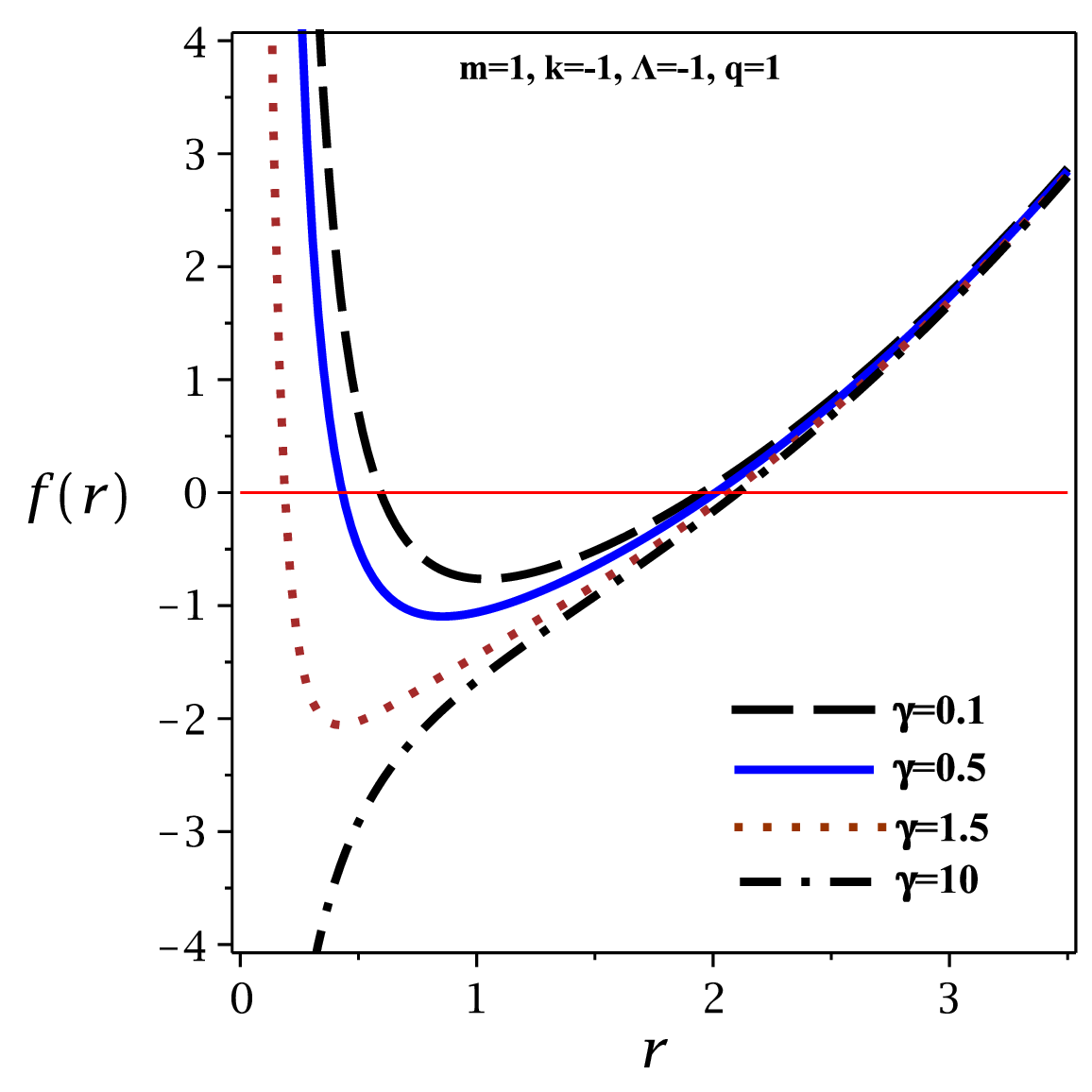} \includegraphics[width=0.35%
\linewidth]{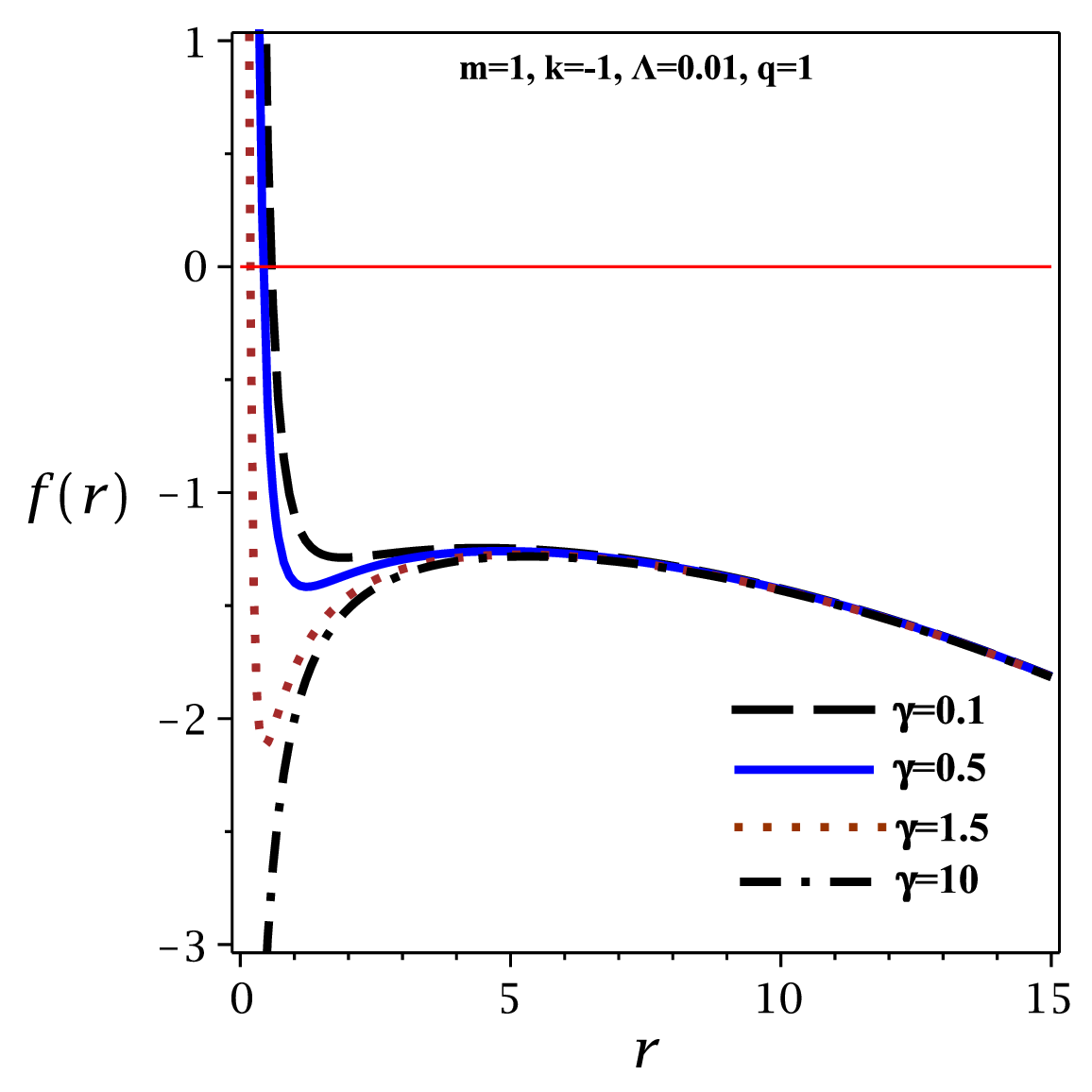} \newline
\caption{The metric function $f(r)$ versus $r$ for $k=-1$, and different
values of the ModMax's parameters. Left panel for AdS case ($\Lambda<0$).
Right panel for dS case ($\Lambda>0$).}
\label{Fig3b}
\end{figure}

Here, we want to study the effect of the ModMax's parameter and topological
constant $k$ on the obtained metric function (\ref{f(r)}). For this purpose,
we plot $f\left( r\right) $ versus $r$ in Figs. \ref{Fig1}$-$\ref{Fig3b}.
Our findings are:

i) For $\Lambda <0$ two roots can be found, one of which is related to the
inner horizon and the other to the event horizon. It is noteworthy that the
number of roots decreases from two to one when $\gamma$ is increased (see
left panels in Figs. \ref{Fig1}$-$\ref{Fig3b}). As one can see in the left
panel of Fig. \ref{Fig1}, the large black holes belong to the negative value
of the topological constant, i.e. $k=-1$.

ii) For $\Lambda >0$ we can have three, two, and one root(s), see the right
panels in Figs. \ref{Fig1}$-$\ref{Fig3b}. Our results in the left panel of
Fig. \ref{Fig1} show that for $k=0$, $k=-1$, there is no event horizon.
However, for $k=+1$, there may be three roots comprising an inner horizon
(the small root), an event horizon (the middle root), and a cosmological
horizon (the largest root), see the right panels in Figs. \ref{Fig1} and \ref%
{Fig2}. In addition, by increasing $\gamma$, the number of roots also
changes.

%%%%%%%%%%%%%%%%%%%%%%%%%%%%%%%%%%%%%%%%%%%%%%%%%%%%%%%%%%%%%%%
%%%%%%%%%%%%%%%%%%%%%%%%%%%%%%%%%%%%%%%%%%%%%%%%%%%%%%%%%%%%%%%
%%%%%%%%%%%%%%%%%%%%%%%%%%%%%%%%%%%%%%%%%%%%%%%%%%%%%%%%%%%%%%%
%%%%%%%%%%%%%%%%%%%%%%%%%%%%%%%%%%%%%%%%%%%%%%%%%%%%%%%%%%%%%%%
%%%%%%%%%%%%%%%%%%%%%%%%%%%%%%%%%%%%%%%%%%%%%%%%%%%%%%%%%%%%%%
%%%%%%%%%%%%%%%%%%%%%%%%%%%%%%%%%%%%%%%%%%%%%%%%%%%%%%%%%%%%%%%$
%%%%%%%%%%%%%%%%%%%%%%%%%%%%%%%%%%%%%%%%%%%%%%%%%%%%%%%%%%%%%%%
%%%%%%%%%%%%%%%%%%%%%%%%%%%%%%%%%%%%%%%%%%%%%%%%%%%%%%%%%%%%%%%

\section{Conserved and Thermodynamic Quantities: The First Law of
Thermodynamics}

To investigate the thermodynamic properties of these black holes, we first
obtain the Hawking temperature of the black holes. For this purpose, we
express the geometrical mass ($m$) in terms of the radius of the event
horizon ($r_{+}$), the cosmological constant ($\Lambda $) and the electrical
charge ($q$), the ModMax parameter ($\gamma$) and the topological constant ($%
k$). We therefore solve $g_{tt}=f(r)=0$, which leads to the following result 
\begin{equation}
m=kr_{+}-\frac{\Lambda r_{+}^{3}}{3}+\frac{q^{2}e^{-\gamma }}{r_{+}}.
\label{mm}
\end{equation}

The Hawking temperature is defined as 
\begin{equation}
T_{H}=\frac{\kappa }{2\pi },  \label{TH1}
\end{equation}
where $\kappa $ is related to the surface gravity by 
\begin{equation}
\kappa =\sqrt{\frac{-1}{2}\left( \nabla _{\mu }\chi _{\nu }\right) \left(
\nabla ^{\mu }\chi ^{\nu}\right)}.  \label{kk}
\end{equation}

By using the metric (\ref{Metric}), the Killing vector $\chi =\partial _{t}$%
, and Eq. (\ref{kk}), we can extract the surface gravity which leads to 
\begin{equation}
\kappa =\frac{1}{2}\left. \frac{\partial f(r)}{\partial r}\right\vert
_{r=r_{+}},  \label{kk2}
\end{equation}%
so, by applying Eqs. (\ref{mm})-(\ref{kk2}), we can get the Hawking
temperature in the following form 
\begin{equation}
T_{H}=\frac{1}{4\pi }\left( \frac{k}{r_{+}}-\Lambda r_{+}-\frac{%
q^{2}e^{-\gamma }}{r_{+}^{3}}\right) ,  \label{TemII}
\end{equation}%
where it depends on the cosmological constant ($\Lambda $), electrical
charge ($q$), the ModMax's parameter ($\gamma $), and topological constant ($%
k$). Furthermore, the Hawking temperature of small black holes (or in high
energy limit of the Hawking temperature) depends on the ModMax's parameter
and electrical charge, i.e. 
\begin{equation}
\underset{r_{+}\rightarrow 0}{\lim }T\propto -\frac{q^{2}e^{-\gamma }}{4\pi
r_{+}^{3}},
\end{equation}%
but for large black holes (or the asymptotic limit of the Hawking
temperature) it only depends on the cosmological constant as 
\begin{equation}
\underset{r_{+}\rightarrow 0}{\lim }T\propto -\frac{\Lambda r_{+}}{4\pi }.
\end{equation}

Using the Gauss law, we can extract the electric charge of black hole per
unit volume ($\mathcal{V}$), in the following form 
\begin{equation}
Q=\frac{\widetilde{Q}}{\mathcal{V}}=\frac{F_{tr}}{4\pi }\int_{0}^{2\pi
}\int_{0}^{\pi }\sqrt{g_{k}}d\theta d\varphi =\frac{q}{4\pi }.  \label{Q}
\end{equation}%
where $F_{tr}=\frac{q}{r^{2}}$, and for case $t=$ constant and $r=$constant,
the determinant of metric tensor $g_{k}$ is $r^{4}\det \left( d\Omega
_{k}^{2}\right) $ (i.e., $g_{k}=\det \left( g_{k}\right) =r^{4}\det \left(
d\Omega _{k}^{2}\right) $). Furthermore, $\mathcal{V}=\int_{0}^{2\pi
}\int_{0}^{\pi }\sqrt{\det \left( d\Omega _{k}^{2}\right) }d\theta d\varphi $%
, where is the area of a unit volume of constant ($t$, $r$) space. For
example, $\mathcal{V}$ is $4\pi $ for $k=1$.

The electric potential at the event horizon ($U$) with respect to the
reference ($r\rightarrow \infty $) is obtained as 
\begin{equation}
U=A_{\mu }\chi ^{\mu }\left\vert _{r\rightarrow \infty }\right. -A_{\mu
}\chi ^{\mu }\left\vert _{r\rightarrow r_{+}}\right. =\frac{qe^{-\gamma }}{%
r_{+}},  \label{elcpo}
\end{equation}%
where the gauge potential is zero when $r\rightarrow \infty $.

Applying the area law, we can get the entropy of the topological ModMax
black holes per unit volume ($\mathcal{V}$), which leads to 
\begin{equation}
S=\frac{\widetilde{S}}{\mathcal{V}}=\frac{\mathcal{A}}{4\mathcal{V}}=\frac{%
\left. \int_{0}^{2\pi }\int_{0}^{\pi }\sqrt{g_{\theta \theta }g_{\varphi
\varphi }}\right\vert _{r=r_{+}}}{4}=\frac{r_{+}^{2}}{4},  \label{S}
\end{equation}%
where $\mathcal{A}$ is the horizon area.

Using Ashtekar-Magnon-Das (AMD) approach \cite{AMDI,AMDII}, we can get the
total mass of the topological ModMax black holes per unit volume ($\mathcal{V%
}$), which is 
\begin{equation}
M=\frac{\widetilde{M}}{\mathcal{V}}=\frac{m}{8\pi }=\frac{1}{8\pi }\left(
kr_{+}-\frac{\Lambda r_{+}^{3}}{3}+\frac{q^{2}e^{-\gamma }}{r_{+}}\right) ,
\label{MM}
\end{equation}%
in the above equation, we use from the geometrical mass (\ref{mm}).

Now, we are in a position to indicate that the obtained conserved and
thermodynamics quantities in Eqs. (\ref{TemII}), (\ref{Q}), (\ref{elcpo}), (%
\ref{S}), and (\ref{MM}), satisfy the first law of thermodynamics in the
following form 
\begin{equation}
dM=TdS+UdQ,
\end{equation}%
where $T=\left( \frac{\partial M}{\partial S}\right) _{Q}$, and $U=\left( 
\frac{\partial M}{\partial Q}\right) _{S}$ are, respectively, in agreement
with those of calculated in Eqs. (\ref{TemII}) and (\ref{elcpo}).

\section{Quasinormal Modes}

As already mentioned, the QNMs represent the characteristic frequencies at
which the scalar field oscillates and the rate at which it is damped. The
QNMs provide a crucial insight into the ModMax parameter's influence on the
scalar field's behavior and its impact on the effective potential. In order
to determine the corresponding QNMs, we need to solve a Schrodinger-like
equation while satisfying certain boundary conditions. These conditions
include the requirement for purely incident waves at the event horizon and
purely outgoing waves at infinity. Due to the inherent complexity of the
differential equation, it is often not possible to find an exact analytical
solution. Analytical solutions for QNMs are only possible under certain
conditions. To overcome this challenge, we rely on a variety of
semi-analytical and numerical methods to obtain the aforementioned
frequencies. For potential barriers that exhibit well-behaved
characteristics, the semi-analytical approach of WKB is a valuable method
for finding QNMs.

\subsection{Spin zero case (scalar field)}

The corresponding wave equation of a massless minimally coupled scalar field 
$\Phi$ is described by the Klein-Gordon equation, namely \cite%
{coupling,p11,p12} 
\begin{equation}
\frac{1}{\sqrt{-g}} \partial_\mu \Bigl( \sqrt{-g} g^{\mu \nu} \partial_\nu %
\Bigl) \Phi = 0.
\end{equation}

To progress and obtain the QNMs, it is convenient to apply the method of
separation of variables, which is the conventional approach, i.e. 
\begin{equation}
\Phi (t,r,\theta ,\phi )=e^{-i\omega t}\frac{\psi (r)}{r}Y_{\ell}^{m}(\theta
,\phi ),  \label{separable}
\end{equation}%
where $Y_{\ell}^{m}$ are the spherical harmonics and $\omega $ is the QN
frequency to be determined. Thus, by applying the above ansatz, we can
decouple the differential equation to obtain an ordinary differential
equation for the radial part only, which is a Schr{\"{o}}dinger-like
equation of the following form 
\begin{equation}
\frac{\mathrm{d}^{2}\psi }{\mathrm{d}x^{2}}+[\omega ^{2}-V(x)]\psi =0,
\end{equation}%
in which $x$ is the so-called tortoise coordinate, defined using the
following expression 
\begin{equation}
x\equiv \int \frac{\mathrm{d}r}{f(r)}.
\end{equation}

In addition, the effective potential barrier can be found with the help of
the following expression 
\begin{equation}
V_{s}(r)=f(r) \left( \frac{\ell (\ell +1)}{r^{2}}+\frac{f^{\prime }(r)}{r}%
\right) ,\;\;\;\;\ell \geq 0
\end{equation}%
where the prime represents differentiation with respect to $r$, and $\ell $
is the angular degree. In particular, for the lapse function given in Eq. %
\eqref{f(r)}, the effective potential takes the simple form 
\begin{equation}
V_{s}(r)=\left( k-\frac{m}{r}+\frac{q^{2}e^{-\gamma }}{r^{2}}-\frac{\Lambda
r^{2}}{3}\right) \Bigg[\frac{\ell (\ell +1)}{r^{2}}+\frac{1}{r}\left( 
\frac{m}{r^{2}}-\frac{2q^{2}e^{-\gamma }}{r^{3}}-\frac{2\Lambda r}{3}\right) %
\Bigg].  \label{Vss}
\end{equation}%
%
%%%%%%%%%%%%%%%%%%%%%%%%%%%%%%%%%%%%%%%%%%%%%%%%%%%%%%%%%%%%%%%%%%%%%%%%%%% 

At this point, some comments are in order. First, from the left column of
Figure \eqref{Fig4} we observe the effect of the ModMax parameter $\gamma$
on the effective potential for different values of the angular number $\ell$
for scalar perturbations.

We observe that, starting from $\gamma=0$ and for increasing values, the
maximum of the effective potential is shifted to the right and it also
decreases. In addition, starting from $\ell=6$ and for increasing values of $%
\ell$, the maximum of the potential increases (comparing different values of 
$\ell$). It should be noted that the ModMax parameter affects the QN
frequencies in a non-trivial way and we can read its effect directly from
the scalar effective potential (see Eq. \eqref{Vss} for details). Since
the ModMax parameter only modifies the charged term, we have two clearly
defined extremal cases: i) when $\gamma \rightarrow 0$ we recover a charged
(Reissner-Nordstrom SdS or SAdS) black hole, and ii) when $\gamma
\rightarrow \infty$ we recover the pure Schwarzschild-de Sitter (SdS) or
Schwarzschild-Anti-de Sitter (SAdS) black hole. According to our results
(see Table \eqref{table:First_set} for details), we can ensure that:

\begin{itemize}
\item When the ModMax parameter $\gamma$ goes from zero to positive values,
the quasinormal modes decrease. To be more precise, fixing $\{ n, \ell \}$,
we notice that as $\gamma$ increases, the QN frequencies also decrease.

\item Fixing the ModMax parameter $\gamma$, we observe that as $n$
increases, the QN frequencies decrease, regardless of the angular degree $%
\ell$.

\item The ModMax parameter $\gamma$ does not alter the stability of the
black solution against scalar perturbations, as confirmed by the minus sign
of the imaginary part of the QN frequencies. This solution is therefore
stable according to this criterion.
\end{itemize}

%%%%%%%%%%%%%%%%%%%%%%%%%%%%%%%%%%%%%%%%%%%%%%%%%%%%%%%%%%%%%%%%%%%%%%%%%
\begin{table*}[tbp]
\caption{QN frequencies for scalar perturbations, setting $k=1$, $m=2$ $%
\Lambda=0.1$, $q=1$ for several different values of the ModMax parameter $%
\protect\gamma$ and varying $\ell$. For comparison reasons, we also show the
frequencies of the classical geometry, i.e., when $\protect\gamma =0$. It
should be noted that as the ModMax parameter increases, the quasinormal
frequencies decrease. Thus, the ModMax parameter has a screening effect on
the QNMs.}\centering
\resizebox{0.95\columnwidth}{!}{\begin{tabular}{ccccccc}
\hline
$\gamma$ & $n$ &  $\ell=6$ & $\ell=7$ & $\ell=8$ & $\ell=9$ & $\ell=10$\\
\hline
0.0 & 0 & 1.105370\, -0.0602992 i & 1.276770\, -0.0603201 i & 1.44803\, -0.0603337 i & 1.61919\, -0.0603429 i & 1.79028\, -0.0603498 i \\
0.0 & 1 & 1.103310\, -0.1809210 i & 1.274790\, -0.1810040 i & 1.44623\, -0.1810430 i & 1.61762\, -0.1810580 i & 1.78886\, -0.1810740 i \\
0.0 & 2 & 1.100060\, -0.3013760 i & 1.270920\, -0.3018030 i & 1.44245\, -0.3019130 i & 1.61444\, -0.3018660 i & 1.78598\, -0.3018750 i \\
0.0 & 3 & 1.098260\, -0.4204130 i & 1.265350\, -0.4227110 i & 1.43618\, -0.4232270 i & 1.60960\, -0.4228450 i & 1.78158\, -0.4228250 i \\
\hline 
0.2 & 0 & 0.928142\, -0.0593617 i & 1.072230\, -0.0593736 i & 1.21618\, -0.0593814 i & 1.36002\, -0.0593870 i & 1.50380\, -0.0593911 i \\
0.2 & 1 & 0.926750\, -0.1781290 i & 1.071010\, -0.1781560 i & 1.21512\, -0.1781690 i & 1.35907\, -0.1781820 i & 1.50294\, -0.1781900 i \\
0.2 & 2 & 0.923966\, -0.2970300 i & 1.068550\, -0.2970520 i & 1.21305\, -0.2970210 i & 1.35716\, -0.2970400 i & 1.50122\, -0.2970400 i \\
0.2 & 3 & 0.919796\, -0.4161530 i & 1.064730\, -0.4161860 i & 1.21006\, -0.4159410 i & 1.35427\, -0.4160100 i & 1.49864\, -0.4159740 i \\
\hline 
0.5 & 0 & 0.770472\, -0.0535483 i & 0.890189\, -0.0535569 i & 1.00977\, -0.0535626 i & 1.12927\, -0.0535666 i & 1.24871\, -0.0535695 i \\
0.5 & 1 & 0.769382\, -0.1606750 i & 0.889241\, -0.1606940 i & 1.00894\, -0.1607060 i & 1.12852\, -0.1607150 i & 1.24803\, -0.1607210 i \\
0.5 & 2 & 0.767213\, -0.2678890 i & 0.887352\, -0.2679000 i & 1.00725\, -0.2679080 i & 1.12702\, -0.2679090 i & 1.24668\, -0.2679080 i \\
0.5 & 3 & 0.764006\, -0.3752250 i & 0.884536\, -0.3752120 i & 1.00470\, -0.3752150 i & 1.12474\, -0.3751860 i & 1.24465\, -0.3751530 i \\
\hline 
0.7 & 0 & 0.698603\, -0.0499479 i & 0.807188\, -0.0499551 i & 0.915651\, -0.0499599 i & 1.02403\, -0.0499633 i & 1.13235\, -0.0499658 i \\
0.7 & 1 & 0.697670\, -0.1498670 i & 0.806374\, -0.1498840 i & 0.914935\, -0.1498940 i & 1.02339\, -0.1499010 i & 1.13177\, -0.1499060 i \\
0.7 & 2 & 0.695814\, -0.2498500 i & 0.804739\, -0.2498690 i & 0.913498\, -0.2498700 i & 1.02211\, -0.2498730 i & 1.13061\, -0.2498750 i \\
0.7 & 3 & 0.693065\, -0.3499210 i & 0.802259\, -0.3499590 i & 0.911336\, -0.3499190 i & 1.02017\, -0.3499040 i & 1.12886\, -0.3498910 i \\
\hline 
1.0 & 0 & 0.618864\, -0.045429 i & 0.715086\, -0.0454346 i & 0.811196\, -0.0454385 i & 0.907230\, -0.0454412 i & 1.00321\, -0.0454431 i \\
1.0 & 1 & 0.618111\, -0.136304 i & 0.714445\, -0.1363150 i & 0.810626\, -0.1363240 i & 0.906717\, -0.1363310 i & 1.00275\, -0.1363360 i \\
1.0 & 2 & 0.616586\, -0.227237 i & 0.713183\, -0.2272190 i & 0.809494\, -0.2272350 i & 0.905685\, -0.2272450 i & 1.00181\, -0.2272480 i \\
1.0 & 3 & 0.614238\, -0.318292 i & 0.711364\, -0.3181300 i & 0.807822\, -0.3181750 i & 0.904120\, -0.3182060 i & 1.00039\, -0.3181990 i \\
\hline 
1.5 & 0 & 0.531635\, -0.0399653 i & 0.614316\, -0.0399695 i & 0.696901\, -0.0399722 i & 0.779418\, -0.0399741 i & 0.861887\, -0.0399755 i \\
1.5 & 1 & 0.531082\, -0.1199080 i & 0.613832\, -0.1199180 i & 0.696486\, -0.1199220 i & 0.779043\, -0.1199270 i & 0.861552\, -0.1199300 i \\
1.5 & 2 & 0.529938\, -0.1998980 i & 0.612808\, -0.1999130 i & 0.695650\, -0.1998900 i & 0.778279\, -0.1998990 i & 0.860884\, -0.1998940 i \\
1.5 & 3 & 0.528097\, -0.2800320 i & 0.611085\, -0.2800660 i & 0.694380\, -0.2798920 i & 0.777082\, -0.2799180 i & 0.859888\, -0.2798720 i \\
\hline 
2.0 & 0 & 0.478469\, -0.0364115 i & 0.552891\, -0.0364148 i & 0.627224\, -0.0364171 i & 0.701496\, -0.0364186 i & 0.775725\, -0.0364197 i \\
2.0 & 1 & 0.478071\, -0.1092340 i & 0.552536\, -0.1092470 i & 0.626898\, -0.1092550 i & 0.701208\, -0.1092580 i & 0.775461\, -0.1092620 i \\
2.0 & 2 & 0.477360\, -0.1820240 i & 0.551870\, -0.1820700 i & 0.626243\, -0.1821050 i & 0.700635\, -0.1821040 i & 0.774926\, -0.1821120 i \\
2.0 & 3 & 0.476573\, -0.2546100 i & 0.551019\, -0.2548120 i & 0.625242\, -0.2549820 i & 0.699793\, -0.2549540 i & 0.774097\, -0.2549860 i \\
\hline 
\end{tabular}
\label{table:First_set}
}
\end{table*}

%%%%%%%%%%%%%%%%%%%%%%%%%%%%%%%%%%%%%%%%%%%%%%%%%%%%

\subsection{Spin one case (Maxwell field)}

%%%%%%%%%%%%%%%%%%%%%%%%%%%%%%%%%%%%%%%%%%%%%%%%%%%%

In this section, we will introduce electromagnetic perturbations. It is
essential to point out that assuming the ModMax non-linear electrodynamics,
the effective Maxwell's equations take the particular form: 
\begin{equation}
\partial _{\mu }\Bigl( \sqrt{-g}e^{-\gamma }F^{\mu \nu }\Bigl) =0.
\end{equation}

Notice that the last equation produces modified Maxwell's equations only if $%
\gamma$ is a function of the spacetime. In this case, we take $\gamma$ as a
constant value, which means that the Maxwell's equations are not modified.
So, the electromagnetic perturbations are are dictated by the standard
Maxwell's equations \cite{Cardoso:2001bb} according to 
\begin{equation}
\nabla _{\nu }F^{\mu \nu }=0,
\end{equation}
where $\nabla _{\nu }$ denotes covariant derivative.

In light that we are considering a spherically symmetric background
geometry, it is possible to expand the corresponding Maxwell potential in
spherical harmonics (for further details see \cite{Cardoso:2001bb} and
references therein). To study electromagnetic perturbations, it is
convenient to use the well-known method of separation of variables to reduce
the situation to a one-dimensional problem. More precisely, after splitting
the radial and angular parts, we obtain a reduced Schr{\"o}dinger-like
equation with a slightly different effective potential for electromagnetic
perturbations. Specifically, the potential takes the reduced form \cite%
{Cardoso:2003sw,Cardoso:2001bb} 
\begin{equation}
V_{EM}(r) = f(r) \: \left( \frac{\ell (\ell+1)}{r^2} \right), \; \; \; \;
\ell \geq 1
\end{equation}
at this point, it becomes evident that to obtain the QNMs for a given
geometry, we only need to specify the lapse function $f(r)$ and subsequently
solve the well-known Regge-Wheeler equation 
\begin{equation}
\frac{\mathrm{d}^2 \psi}{\mathrm{d}x^2} + [ \omega^2 - V_{EM}(x) ] \psi = 0.
\end{equation}

The concrete form of the effective potential for electromagnetic
perturbations is given as follow: 
\begin{equation}
V_{EM}(r)=\left( k-\frac{m}{r}+\frac{q^{2}e^{-\gamma }}{r^{2}}-\frac{\Lambda
r^{2}}{3}\right) \Bigg(\frac{\ell (\ell +1)}{r^{2}}\Bigg).  \label{VEM}
\end{equation}

We show, in figures, the effective potential for electromagnetic
perturbations in Figure \eqref{Fig4}, middle panel, setting the parameters $%
\{ k, m, \Lambda, q \}$ and varying $\ell$ from $6$ to $9$ as well as the
ModMax parameter $\gamma$ from $0$ to $2$, see Eq. \eqref{VEM} for
explicit form. We observe that, similar to the massless scalar case, in the
electromagnetic case the maximum of the potential increases when the angular
number $\ell$ increases. Similarly, the maximum is shifted to the right when 
$\ell$ increases too. Even more, for a fixed $\ell$, we vary the ModMax
parameter $\gamma$ and we observe that increasing it the effective potential
decreases. Such effect transform the effective potential from a
Reissner-Nordstrom SdS or SAdS black hole to a pure Schwarzschild-de Sitter
(SdS)or Schwarzschild-Antide-Sitter (SAdS) black hole. In light of our
results (see Table \eqref{table:Second_set} for details), we can confirm
that:

\begin{itemize}
\item When the ModMax parameter $\gamma$ goes from zero to positive values,
the quasinormal modes decrease, in other words, fixing $\{ n, \ell \}$, we
notice that as $\gamma$ increases, the QN frequencies also decrease.

\item Fixing the ModMax parameter $\gamma$, we observe that as $n$
increases, the QN frequencies decrease, regardless of the angular degree $%
\ell$.

\item As occurs in the massless scalar case, the ModMax parameter $\gamma$
does not modify the stability of the black solution against electromagnetic
perturbations, as is revealed by the minus sign of the imaginary part of the
QN frequencies. Thus, according to this criterion the solution is stable.
\end{itemize}

%%%%%%%%%%%%%%%%%%%%%%%%%%%%%%%%%%%%%%%%%%%%%%%%%%%
%%%%%%%%%%%%%%%%%%%%%%%%%%%%%%%%%%%%%%%%%%%%%%%%%%%%
\begin{table*}[tbp]
\caption{QN frequencies for electromagnetic perturbations, setting $k=1$, $%
m=2$ $\Lambda=0.1$, $q=1$ for several different values of the ModMax
parameter $\protect\gamma$ and varying $\ell$. For comparison reasons, we
also show the frequencies of the classical geometry, i.e., when $\protect%
\gamma =0$. As in the scalar case, when the ModMax parameter increases, the
quasinormal frequencies decrease. Thus, the ModMax parameter has a screening
effect on the QNMs.}\centering
\resizebox{0.95\columnwidth}{!}{\begin{tabular}{ccccccc}
\hline
$\gamma$ & $n$ &  $\ell=6$ & $\ell=7$ & $\ell=8$ & $\ell=9$ & $\ell=10$\\
\hline
0.0 & 0 & 1.10455\, -0.060301 i & 1.27611\, -0.0603194 i & 1.44744\, -0.0603333 i & 1.61866\, -0.060343 i & 1.78981\, -0.0603497 i \\
0.0 & 1 & 1.10197\, -0.181011 i & 1.27422\, -0.1809900 i & 1.44569\, -0.1810350 i & 1.61702\, -0.181066 i & 1.78838\, -0.1810740 i \\
0.0 & 2 & 1.09572\, -0.302344 i & 1.27085\, -0.3016570 i & 1.44223\, -0.3018340 i & 1.61346\, -0.301949 i & 1.78548\, -0.3018760 i \\
0.0 & 3 & 1.08260\, -0.426147 i & 1.26725\, -0.4218510 i & 1.43715\, -0.4227600 i & 1.60713\, -0.423353 i & 1.78102\, -0.4228370 i \\
\hline 
0.2 & 0 & 0.927681\, -0.0593616 i & 1.07183\, -0.0593734 i & 1.21583\, -0.0593813 i & 1.35971\, -0.059387 i & 1.50352\, -0.059391 i \\
0.2 & 1 & 0.926250\, -0.1781360 i & 1.07061\, -0.1781560 i & 1.21477\, -0.1781690 i & 1.35876\, -0.178182 i & 1.50266\, -0.178190 i \\
0.2 & 2 & 0.923251\, -0.2971100 i & 1.06812\, -0.2970590 i & 1.21268\, -0.2970230 i & 1.35685\, -0.297039 i & 1.50094\, -0.297038 i \\
0.2 & 3 & 0.918293\, -0.4166210 i & 1.06420\, -0.4162310 i & 1.20965\, -0.4159570 i & 1.35398\, -0.416003 i & 1.49839\, -0.415962 i \\
\hline
0.5 & 0 & 0.770210\, -0.0535481 i & 0.889961\, -0.0535568 i & 1.00958\, -0.0535625 i & 1.12909\, -0.0535666 i & 1.24855\, -0.0535695 i \\
0.5 & 1 & 0.769126\, -0.1606730 i & 0.889010\, -0.1606940 i & 1.00874\, -0.1607060 i & 1.12834\, -0.1607140 i & 1.24787\, -0.1607200 i \\
0.5 & 2 & 0.766992\, -0.2678730 i & 0.887102\, -0.2679050 i & 1.00706\, -0.2679060 i & 1.12684\, -0.2679080 i & 1.24651\, -0.2679070 i \\
0.5 & 3 & 0.763917\, -0.3751350 i & 0.884220\, -0.3752460 i & 1.00453\, -0.3752030 i & 1.12458\, -0.3751780 i & 1.24449\, -0.3751530 i \\
\hline
0.7 & 0 & 0.698406\, -0.0499479 i & 0.807018\, -0.0499551 i & 0.915502\, -0.0499599 i & 1.02390\, -0.0499633 i & 1.13223\, -0.0499657 i \\
0.7 & 1 & 0.697459\, -0.1498700 i & 0.806198\, -0.1498850 i & 0.914785\, -0.1498940 i & 1.02326\, -0.1499010 i & 1.13165\, -0.1499060 i \\
0.7 & 2 & 0.695528\, -0.2498820 i & 0.804528\, -0.2498810 i & 0.913349\, -0.2498700 i & 1.02197\, -0.2498720 i & 1.13049\, -0.2498740 i \\
0.7 & 3 & 0.692513\, -0.3501000 i & 0.801924\, -0.3500290 i & 0.911187\, -0.3499190 i & 1.02004\, -0.3499010 i & 1.12875\, -0.3498880 i \\
\hline
1.0 & 0 & 0.618727\, -0.045429 i & 0.714968\, -0.0454346 i & 0.811092\, -0.0454385 i & 0.907137\, -0.0454411 i & 1.00313\, -0.0454431 i \\
1.0 & 1 & 0.617968\, -0.136305 i & 0.714319\, -0.1363160 i & 0.810520\, -0.1363250 i & 0.906623\, -0.1363310 i & 1.00266\, -0.1363350 i \\
1.0 & 2 & 0.616408\, -0.227251 i & 0.713020\, -0.2272330 i & 0.809377\, -0.2272380 i & 0.905587\, -0.2272470 i & 1.00173\, -0.2272470 i \\
1.0 & 3 & 0.613935\, -0.318376 i & 0.711067\, -0.3182110 i & 0.807663\, -0.3181960 i & 0.904004\, -0.3182130 i & 1.00034\, -0.3181900 i \\
\hline
1.5 & 0 & 0.531548\, -0.0399653 i & 0.614243\, -0.0399694 i & 0.696836\, -0.0399721 i & 0.779359\, -0.0399741 i & 0.861834\, -0.0399755 i \\
1.5 & 1 & 0.530992\, -0.1199080 i & 0.613770\, -0.1199160 i & 0.696422\, -0.1199220 i & 0.778989\, -0.1199260 i & 0.861499\, -0.1199300 i \\
1.5 & 2 & 0.529830\, -0.1999070 i & 0.612809\, -0.1998890 i & 0.695596\, -0.1998860 i & 0.778247\, -0.1998920 i & 0.860829\, -0.1998940 i \\
1.5 & 3 & 0.527926\, -0.2800790 i & 0.611320\, -0.2799260 i & 0.694363\, -0.2798730 i & 0.777135\, -0.2798770 i & 0.859828\, -0.2798740 i \\
\hline
2.0 & 0 & 0.478405\, -0.0364115 i & 0.552835\, -0.0364149 i & 0.627176\, -0.0364171 i & 0.701453\, -0.0364186 i & 0.775687\, -0.0364197 i \\
2.0 & 1 & 0.478000\, -0.1092360 i & 0.552463\, -0.1092500 i & 0.626850\, -0.1092550 i & 0.701163\, -0.1092590 i & 0.775423\, -0.1092620 i \\
2.0 & 2 & 0.477255\, -0.1820390 i & 0.551704\, -0.1821070 i & 0.626195\, -0.1821050 i & 0.700582\, -0.1821070 i & 0.774890\, -0.1821120 i \\
2.0 & 3 & 0.476352\, -0.2546940 i & 0.550515\, -0.2550200 i & 0.625196\, -0.2549800 i & 0.699707\, -0.2549710 i & 0.774073\, -0.2549830 i \\
\hline
\end{tabular}
\label{table:Second_set}
}
\end{table*}

\subsection{Spin one-half case (Dirac field)}

To study Dirac perturbations (i.e., fermions) we first should remember the
vierbein formalism and, subsequently, we will review and show the wave
equation and the effective potential (see also \cite{Destounis:2018qnb} for
more details) the tetrad (or vierbein), $e_a^\mu$, is defined as follows 
\begin{equation}
e_\mu^a e_\nu^b \eta_{ab} = g_{\mu \nu},
\end{equation}
as is usual, $\eta_{ab}$ is the flat Minkowski metric tensor. Also, the
vierbein encodes two different kinds of indices: i) a flat index $a$, and
ii) a spacetime index $\mu$, and it may be viewed as the "square root" of
the metric tensor $g_{\mu \nu}$.

The Dirac matrices in a curved spacetime are given as follows 
\begin{equation}
G^{\mu }\equiv e_{a}^{\mu }\gamma ^{a},
\end{equation}%
and they satisfy the following property 
\begin{eqnarray}
\{\gamma ^{a},\gamma ^{b}\} &=&-2\eta ^{ab}, \\
&&  \notag \\
\{G^{\mu },G^{\nu }\} &=&-2g^{ab},
\end{eqnarray}%
where $\gamma ^{a}$ are the Dirac matrices used in relativistic quantum
mechanics (in flat spacetime).

Last but not least, let us introduce the spin connection $\omega _{ab\mu
},\Gamma _{\mu }$ defined as follows 
\begin{eqnarray}
\Gamma _{\mu } &=&-\frac{1}{8}\omega _{ab\mu }[\gamma ^{a},\gamma ^{b}], \\
&&  \notag \\
\omega _{ab\mu } &=&\eta _{ac}[e_{\nu }^{c}e_{b}^{\lambda }\Gamma _{\mu
\lambda }^{\nu }-e_{b}^{\lambda }\partial _{\mu }e_{\lambda }^{c}],
\end{eqnarray}%
where $\Gamma _{\mu \lambda }^{\nu }$ are the Christoffel symbols.

At this point, it becomes natural to introduce the equation responsible for
describing perturbation for fermions, i.e., the Dirac equation. Let us
consider a spin one-half fermion $\Psi $ in curved spacetime, in that case,
the Dirac equation takes the form 
\begin{equation}
(iG^{\mu }D_{\mu }-m_{f})\Psi =0,
\end{equation}%
being the mass of the fermion $m_{f}$, and the covariant derivative defined
as $D_{\mu }\equiv \partial _{\mu }+\Gamma _{\mu }$. The process is
equivalent to the scalar and Maxwell case, namely, we should use the method
of separation of variables, where now we use the spinor spherical harmonics 
\cite{Finster:1998ak}, and two radial parts 
\begin{align}
& r^{-1} f(r)^{-1/4} F(r), \\
&  \notag \\
& r^{-1} f(r)^{-1/4} iG(r),
\end{align}%
for the upper and lower components of the Dirac spinor $\Psi $,
respectively. For the massless case ($m_{f}=0$), the equation is reduced to
be \cite{Destounis:2018qnb} 
\begin{eqnarray}
\frac{dF}{dx}-WF+\omega G &=&0,  \label{FF} \\
&&  \notag \\
\frac{dG}{dx}+WG-\omega F &=&0.  \label{GG}
\end{eqnarray}

As in the previous cases, $x$ is the tortoise coordinate and the function $W$
takes the form \cite{Destounis:2018qnb} 
\begin{equation}
W\equiv \frac{\xi \sqrt{f}}{r},
\end{equation}%
where $\xi \equiv \pm (j+1/2)=\pm 1,\pm 2,...$, with $j$ being the total
angular momentum, $j=\ell \pm 1/2$ \cite{Destounis:2018qnb}.

To conclude this part, notice first that Eqs. \eqref{FF} and \eqref{GG} are
two first-order differential equations for $F$, and $G$. They can be
conveniently combined to obtain a wave equation for each one of the
following form 
\begin{eqnarray}
\frac{d^{2}F}{dx^{2}}+[\omega ^{2}-V_{+}]F &=&0, \\
&&  \notag \\
\frac{d^{2}G}{dx^{2}}+[\omega ^{2}-V_{-}]G &=&0,
\end{eqnarray}%
where the potentials are well-known and are summarized in the following
expression 
\begin{equation}
V_{\pm }=W^{2}\pm \frac{dW}{dx}.
\end{equation}

According to \cite{Cooper:1994eh}, and taking advantage of the language of
supersymmetry, the potential $V_{\pm}$ are superpartners because they are
derived from a superpotential. The natural consequence of that is that they
share the same spectra. Thus, in the following, we shall work with the plus
sign and the wave equation for $F(r)$.

Finally, the effective potential for Dirac perturbations, taking advantage
of the tortoise coordinate and writing it down in terms of $\xi $ is given
as follows 
\begin{equation}
V_{\pm }=f(r)\Bigg(\frac{\xi ^{2}}{r^{2}}\pm \frac{\mathrm{d}}{\mathrm{d}r}%
\left( \frac{\xi }{r}\sqrt{f(r)}\right) \Bigg).  \label{VVp}
\end{equation}%
Taking the positive sign and the concrete form of lapse function we finally
arrive 
\begin{equation}
V_{+}=\frac{\xi \left( 1-\frac{m}{r}+\frac{q^{2}e^{-\gamma }}{r^{2}}-\frac{%
\Lambda r^{2}}{3}\right) \left( \left( \xi \sqrt{1-\frac{m}{r}+\frac{%
q^{2}e^{-\gamma }}{r^{2}}-\frac{\Lambda r^{2}}{3}}+\frac{3m}{2r}-1\right) -%
\frac{2q^{2}e^{-\gamma }}{r^{2}}\right) }{\left( \sqrt{1-\frac{m}{r}+\frac{%
q^{2}e^{-\gamma }}{r^{2}}-\frac{\Lambda r^{2}}{3}}\right) r^{2}}.
\label{VPlus}
\end{equation}

\begin{figure}[tbph]
\centering
\includegraphics[width=0.32\linewidth]{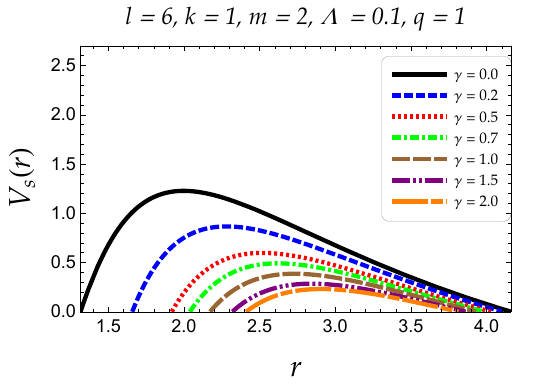} %
\includegraphics[width=0.32\linewidth]{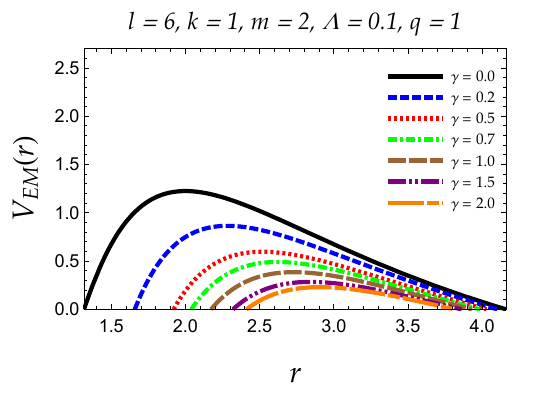} %
\includegraphics[width=0.32\linewidth]{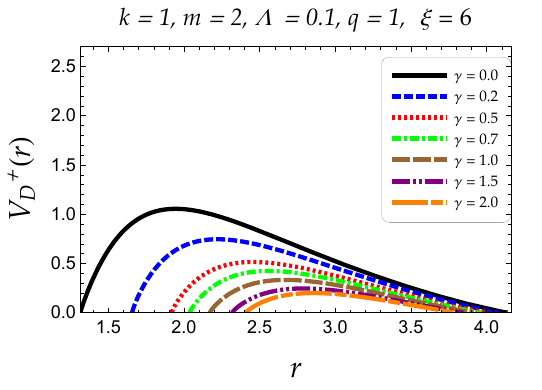} \newline
\includegraphics[width=0.32\linewidth]{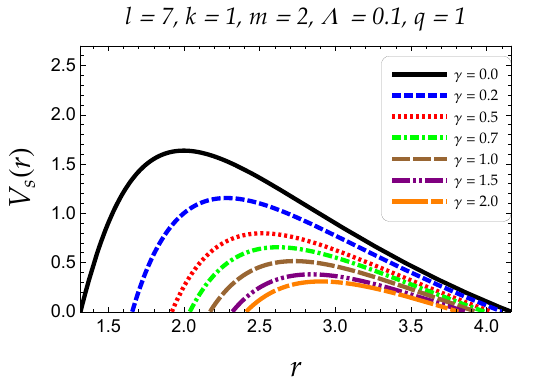} %
\includegraphics[width=0.32\linewidth]{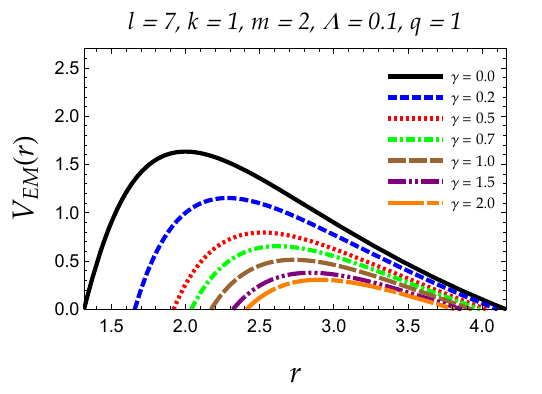} %
\includegraphics[width=0.32\linewidth]{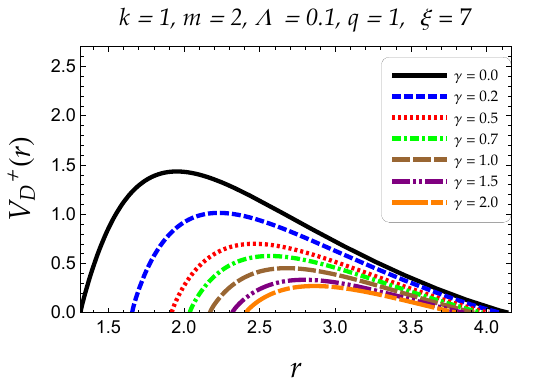} \newline
\includegraphics[width=0.32\linewidth]{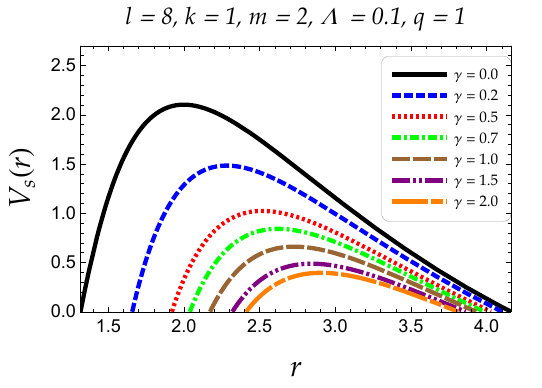} %
\includegraphics[width=0.32\linewidth]{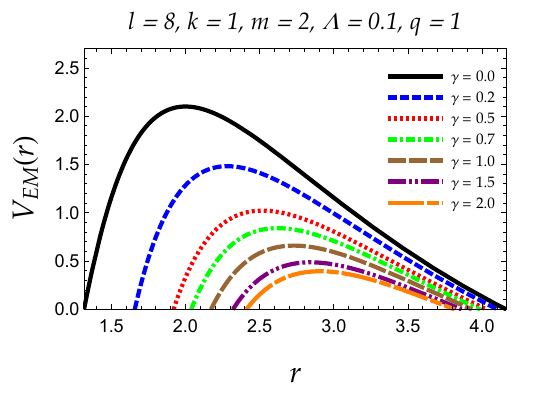} %
\includegraphics[width=0.32\linewidth]{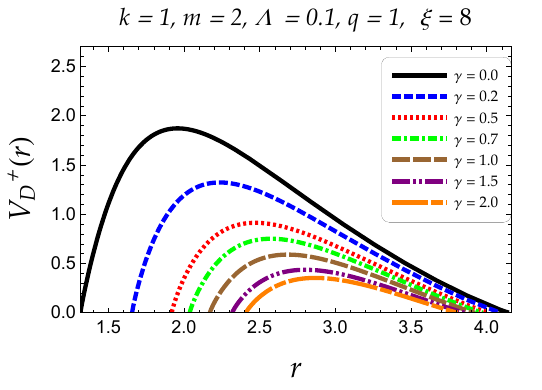} \newline
\includegraphics[width=0.32\linewidth]{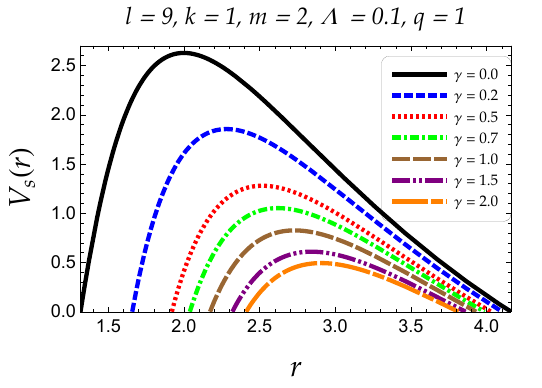} %
\includegraphics[width=0.32\linewidth]{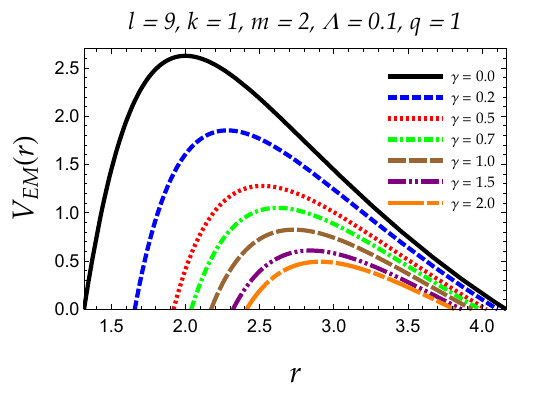} %
\includegraphics[width=0.32\linewidth]{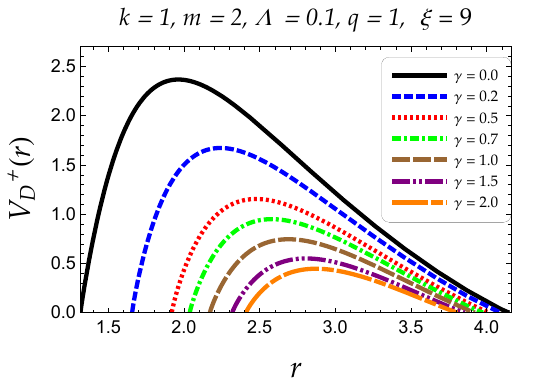} \newline
\caption{ Effective potential for the three types of perturbations
investigated in this paper. \textbf{Left column:} scalar perturbations for
different values of the MaxMod parameter $\protect\gamma$ fixing $\{\ell, k,
m, \Lambda, q\}$. \textbf{Middle column:} electromagnetic perturbations for
different values of the MaxMod parameter $\protect\gamma$ fixing $\{ \ell,
k, m, \Lambda, q\}$. \textbf{Right column:} Dirac perturbations for
different values of the MaxMod parameter $\protect\gamma$ fixing $\{ \protect%
\xi, k, m,\Lambda, q \}$. }
\label{Fig4}
\end{figure}

Finally, let us mention some features which can be observed in Figure %
\eqref{Fig4} right panel, after observe careful the effective potential for
Dirac perturbations, setting the parameters $\{ k, m, \Lambda, q \}$ and
varying $\xi$ from $6$ to $9$ as well as the ModMax parameter $\gamma$ from $%
0$ to $2$, see Eq. \eqref{VPlus} for explicit form. As occurs in the last
two cases, the maximum of the potential increases when the number $\xi$
increases (parameter which is directly related with $\ell$). Also, the
maximum is shifted to the right when $\xi$ increases, equivalently as occur
in the previous two cases. Finally, varying (increasing) the ModMax
parameter $\gamma$ (for a fixed $\xi$), it is observed that the effective
potential decreases.

In light of our results (see Tables \eqref{table:Third_set}, %
\eqref{table:Fourth_set} and \eqref{table:Fifth_set} for details), we can
confirm that:

\begin{itemize}
\item As occur in the previous two cases, when the ModMax parameter $\gamma$
goes from zero to positive values, the quasinormal modes decrease. To
confirm, please check the tables above-mentioned, i.e., fixing $\{ n,
\Lambda, \xi, q \}$, we notice that as $\gamma$ increases, the QN
frequencies also decrease (extracted from Tables \eqref{table:Third_set}, %
\eqref{table:Fourth_set} and \eqref{table:Fifth_set}).

\item Fixing the ModMax parameter $\gamma$, we observe that as $n$
increases, the QN frequencies decrease, regardless of the angular degree $\xi$%
.

\item As occurs in the massless scalar case and electromagnetic case, the
ModMax parameter $\gamma$ does not modify the stability of the black
solution against Dirac perturbations. This statement is confirmed by the
minus sign of the imaginary part of the QN frequencies. Thus, according to
this criterion the solution is stable, at least for the range of parameters
used.
\end{itemize}

%%%%%%%%%%%%%%%%%%%%%%%%%%%%%%%%%%%%%%%%%%%%%%%%%%%%%%%%%%%%%%%
%%%%%%%%%%%%%%%%%%%%%%%%%%%%%%%%%%%%%%%%%%%%%%%%%%%%%%%%%%%%%%%

\begin{table*}[tbp]
\caption{QN frequencies for Dirac perturbations, setting $k=1$, $m=2$ , $%
\protect\xi =5$ and varying the set $\{ \Lambda, q, n \}$ taking the ModMax
parameter $\protect\gamma=0.0$. This case is included for comparison reasons
and it match with the published version using a different method 
\protect\cite{Jing:2003wq}.}\centering
\resizebox{0.95\columnwidth}{!}{\begin{tabular}{ccccccccc}
\hline
$n$ & $\Lambda$ & $\omega(q=0.0)$ &  $\omega(q=0.1)$ & $\omega(q=0.2)$ & $\omega(q=0.3)$ & $\omega(q=0.4)$ & $\omega(q=0.5)$ & $\omega(q=0.6)$\\
\hline
0 & 0.00 & 0.960215\, -0.0962564 i & 0.961827\, -0.0963096 i & 0.966735\, \
-0.0964676 i & 0.975163\, -0.0967242 i & 0.987525\, -0.0970661 i & \
1.0045\, -0.0974654 i & 1.02716\, -0.0978635 i 
\\ 
0 & 0.01 & 0.916158\, \
-0.0918083 i & 0.917846\, -0.0918743 i & 0.922985\, -0.092071 i & \
0.931804\, -0.0923934 i & 0.944726\, -0.0928303 i & 0.962444\, \
-0.0933575 i & 0.986064\, -0.0939224 i 
\\ 
0 & 0.02 & 0.869837\, -0.0871387 i & \
0.871614\, -0.0872189 i & 0.877021\, -0.0874586 i & 0.886293\, \
-0.0878541 i & 0.899863\, -0.0883962 i & 0.918439\, -0.0890645 i & \
0.943151\, -0.0898121 i 
\\ 
0 & 0.03 & 0.820869\, -0.0822089 i & 0.822752\, \
-0.0823053 i & 0.828476\, -0.0825937 i & 0.838281\, -0.0830715 i & \
0.852611\, -0.0837319 i & 0.872186\, -0.0845577 i & 0.898163\, \
-0.0855079 i 
\\ 
0 & 0.04 & 0.768749\, -0.0769683 i & 0.770758\, -0.0770833 i & \
0.776864\, -0.0774278 i & 0.78731\, -0.0780003 i & 0.802547\, \
-0.0787957 i & 0.823308\, -0.0798002 i & 0.850771\, -0.0809784 i 
\\ 
\
0 & 0.05 & 0.712784\, -0.0713475 i & 0.714951\, -0.0714845 i & 0.721528\, \
-0.0718954 i & 0.732763\, -0.0725791 i & 0.749107\, -0.0735323 i & \
0.771305\, -0.0747436 i & 0.800549\, -0.076183 i 
\\ 
0 & 0.06 & 0.651986\, \
-0.0652474 i & 0.654354\, -0.0654113 i & 0.661534\, -0.065903 i & \
0.67377\, -0.0667218 i & 0.69151\, -0.0678646 i & 0.715496\, \
-0.0693218 i & 0.746926\, -0.0710672 i 
\\ 
0 & 0.07 & 0.584848\, -0.0585172 i & \
0.587487\, -0.0587157 i & 0.595475\, -0.0593106 i & 0.609041\, \
-0.0603003 i & 0.628613\, -0.0616801 i & 0.654909\, -0.0634409 i & \
0.689109\, -0.0655559 i 
\\ 
0 & 0.08 & 0.508867\, -0.0509062 i & 0.511899\, \
-0.0511512 i & 0.52105\, -0.0518895 i & 0.536505\, -0.0531081 i & \
0.558629\, -0.054806 i & 0.588067\, -0.0569578 i & 0.625935\, \
-0.0595376 i 
\\ 
0 & 0.09 & 0.419262\, -0.0419362 i & 0.422939\, -0.0422625 i & \
0.433973\, -0.0432053 i & 0.45242\, -0.044782 i & 0.478459\, \
-0.0469606 i & 0.512529\, -0.049621 i & 0.555579\, -0.0528326 i 
\\ 
\
0 & 0.10 & 0.304224\, -0.030426 i & 0.309367\, -0.0318253 i & 0.32402\, \
-0.0303196 i & 0.348472\, -0.0339389 i & 0.381719\, -0.0373302 i & \
0.423659\, -0.0408849 i & 0.474867\, -0.0451985 i 
\\
\hline 
 1 & 0.00 & 0.949593\, -0.290179 i & 0.951226\, -0.290335 i & 0.956197\, -0.2908 i & 0.964738\, -0.291554 i & 0.977273\, -0.292555 i & 0.994493\, -0.293717 i & 1.01751\, -0.29486 i 
 \\
 1 & 0.01 &  0.906891\, -0.27655 i & 0.908593\, -0.276747 i & 0.913774\, -0.277332 i & 0.92267\, -0.27829 i & 0.935712\, -0.279587 i & 0.953608\, -0.281148 i & 0.977488\, -0.28281 i 
 \\
 1 & 0.02 &  0.861871\, -0.262294 i & 0.863655\, -0.262534 i & 0.869085\, -0.263252 i & 0.8784\, -0.264436 i & 0.892041\, -0.266057 i & 0.910729\, -0.268053 i & 0.935617\, -0.270279 i 
 \\
 1 & 0.03 &  0.814144\, -0.247293 i & 0.816028\, -0.247583 i & 0.821756\, -0.248449 i & 0.831574\, -0.249885 i & 0.845929\, -0.251867 i & 0.865557\, -0.254345 i & 0.891632\, -0.257191 i 
 \\
 1 & 0.04 &  0.763201\, -0.231394 i & 0.765206\, -0.23174 i & 0.7713\, -0.232776 i & 0.781729\, -0.234499 i & 0.796951\, -0.236891 i & 0.817709\, -0.239911 i & 0.845199\, -0.24345 i 
 \\
 1 & 0.05 &  0.708342\, -0.214385 i & 0.710499\, -0.214798 i & 0.717051\, -0.216034 i & 0.728244\, -0.218092 i & 0.744537\, -0.220961 i & 0.766683\, -0.224606 i & 0.795889\, -0.228935 i 
 \\
 1 & 0.06 &  0.648571\, -0.195969 i & 0.650926\, -0.196462 i & 0.658068\, -0.197941 i & 0.670241\, -0.200405 i & 0.6879\, -0.203844 i & 0.711791\, -0.208229 i & 0.743127\, -0.21348 i 
 \\
 1 & 0.07 &  0.582371\, -0.175689 i & 0.584995\, -0.176286 i & 0.592935\, -0.178075 i & 0.606423\, -0.181051 i & 0.625888\, -0.1852 i & 0.652056\, -0.190498 i & 0.686114\, -0.196862 i 
 \\
 1 & 0.08 &  0.507228\, -0.152792 i & 0.510239\, -0.153524 i & 0.519341\, -0.155748 i & 0.534704\, -0.159403 i & 0.556711\, -0.164517 i & 0.585994\, -0.170983 i & 0.623682\, -0.178739 i 
 \\
 1 & 0.09 &  0.41834\, -0.125841 i & 0.42201\, -0.12684 i & 0.432969\, -0.12963 i & 0.451341\, -0.134394 i & 0.477297\, -0.141027 i & 0.511125\, -0.148874 i & 0.553987\, -0.15855 i 
 \\
 1 & 0.10 &  0.303869\, -0.0912867 i & 0.310752\, -0.0986849 i & 0.320168\, -0.0839339 i & 0.34697\, -0.0998782 i & 0.3809\, -0.111621 i & 0.422625\, -0.122188 i & 0.473949\, -0.13578 i 
 \\
\hline
 2 & 0.00 & 0.929979\, -0.487634 i & 0.931649\, -0.487888 i & 0.936736\, -0.488639 i & 0.94548\, -0.489855 i & 0.958323\, -0.49146 i & 0.975988\, -0.493304 i & 0.999629\, -0.495074 i 
 \\
 2 & 0.01 & 0.889651\, -0.464109 i & 0.891378\, -0.464433 i & 0.896637\, -0.465397 i & 0.905673\, -0.466974 i & 0.918934\, -0.469103 i & 0.937156\, -0.471652 i & 0.96151\, -0.474339 i 
 \\
 2 & 0.02 & 0.846942\, -0.439644 i & 0.84874\, -0.440045 i & 0.854212\, -0.44124 i & 0.863607\, -0.443208 i & 0.877381\, -0.445901 i & 0.896279\, -0.449207 i & 0.921494\, -0.452877 i 
 \\
 2 & 0.03 & 0.801453\, -0.414041 i & 0.803339\, -0.414525 i & 0.809076\, -0.415975 i & 0.818917\, -0.418376 i & 0.833324\, -0.421688 i & 0.853053\, -0.425822 i & 0.879314\, -0.430559 i 
 \\
 2 & 0.04 & 0.752658\, -0.387036 i & 0.754655\, -0.387616 i & 0.760728\, -0.389354 i & 0.771129\, -0.392241 i & 0.786326\, -0.39625 i & 0.807083\, -0.401305 i & 0.834627\, -0.407223 i 
 \\
 2 & 0.05 & 0.699844\, -0.358275 i & 0.701985\, -0.358966 i & 0.708487\, -0.361041 i & 0.719604\, -0.364491 i & 0.735805\, -0.369301 i & 0.757856\, -0.375409 i & 0.786996\, -0.382662 i 
 \\
  2 & 0.06 & 0.641995\, -0.327251 i & 0.644326\, -0.328078 i & 0.651395\, -0.330557 i & 0.663453\, -0.334687 i & 0.680959\, -0.340449 i & 0.704674\, -0.347796 i & 0.735835\, -0.356596 i 
 \\
 2 & 0.07 & 0.577573\, -0.293201 i & 0.580166\, -0.294201 i & 0.588016\, -0.297194 i & 0.601354\, -0.302173 i & 0.620616\, -0.309116 i & 0.646545\, -0.317994 i & 0.680339\, -0.328655 i 
 \\
  2 & 0.08 & 0.504033\, -0.25486 i & 0.507\, -0.256072 i & 0.516013\, -0.259805 i & 0.531188\, -0.265893 i & 0.552986\, -0.274472 i & 0.581971\, -0.285283 i & 0.619316\, -0.298256 i 
  \\
  2 & 0.09 & 0.416533\, -0.209823 i & 0.420216\, -0.211545 i & 0.430973\, -0.216091 i & 0.449237\, -0.224122 i & 0.475153\, -0.235443 i & 0.508323\, -0.248159 i & 0.550856\, -0.2644 i 
  \\
 2 & 0.10 & 0.30317\, -0.152169 i & 0.317694\, -0.172731 i & 0.303932\, -0.118853 i & 0.341499\, -0.161073 i & 0.378786\, -0.185026 i & 0.419961\, -0.202366 i & 0.47234\, -0.226809 i 
 \\
\hline
3 & 0.00 & 0.903578\, -0.689241 i & 0.905298\, -0.689586 i & 0.910538\, -0.690606 i & 0.919551\, -0.692252 i & 0.932804\, -0.694412 i & 0.951058\, -0.696865 i & 0.975524\, -0.699153 i 
\\
3 & 0.01 &  0.866171\, -0.654985 i & 0.867933\, -0.655434 i & 0.8733\, -0.656771 i & 0.882528\, -0.658954 i & 0.89609\, -0.661892 i & 0.914754\, -0.665393 i & 0.939748\, -0.669046 i 
\\
3 & 0.02 &  0.826383\, -0.619585 i & 0.8282\, -0.620146 i & 0.833735\, -0.62182 i & 0.843246\, -0.624575 i & 0.85721\, -0.628339 i & 0.876406\, -0.632949 i & 0.902075\, -0.638042 i 
\\
3 & 0.03 &  0.78379\, -0.582754 i & 0.785681\, -0.583436 i & 0.791437\, -0.585477 i & 0.801318\, -0.588855 i & 0.815807\, -0.593514 i & 0.835689\, -0.59932 i & 0.86222\, -0.605958 i 
\\
3 & 0.04 &  0.737839\, -0.544118 i & 0.739828\, -0.544936 i & 0.745876\, -0.547388 i & 0.756245\, -0.551459 i & 0.77142\, -0.557111 i & 0.792191\, -0.564234 i & 0.819828\, -0.572563 i 
\\
3 & 0.05 &  0.687788\, -0.503171 i & 0.689906\, -0.504146 i & 0.696346\, -0.507074 i & 0.707363\, -0.51194 i & 0.723446\, -0.518726 i & 0.745378\, -0.527337 i & 0.774443\, -0.537561 i 
 \\
3 & 0.06 &  0.632584\, -0.459197 i & 0.63488\, -0.460361 i & 0.641851\, -0.463854 i & 0.653752\, -0.469675 i & 0.67105\, -0.477797 i & 0.694528\, -0.488151 i & 0.72546\, -0.500556 i 
\\
3 & 0.07 &  0.570648\, -0.411114 i & 0.573199\, -0.412521 i & 0.580919\, -0.41673 i & 0.594048\, -0.423735 i & 0.613024\, -0.433504 i & 0.63863\, -0.446013 i & 0.672066\, -0.461033 i 
\\
3 & 0.08 &  0.499385\, -0.35714 i & 0.502279\, -0.358825 i & 0.51118\, -0.364092 i & 0.526068\, -0.372613 i & 0.547596\, -0.384716 i & 0.576155\, -0.399907 i & 0.613019\, -0.41815 i 
\\
3 & 0.09 &  0.413886\, -0.293896 i & 0.417633\, -0.296394 i & 0.428002\, -0.302599 i & 0.446176\, -0.313984 i & 0.47225\, -0.330248 i & 0.504129\, -0.347491 i & 0.546259\, -0.370413 i 
\\
3 & 0.10 &  0.30214\, -0.213074 i & 0.334279\, -0.253476 i & 0.260522\, -0.124274 i & 0.32869\, -0.21671 i & 0.374779\, -0.257461 i & 0.414903\, -0.281298 i & 0.470328\, -0.31834 i 
\\
\hline
\end{tabular}
\label{table:Third_set}
}
\end{table*}

\begin{table*}[tbp]
\caption{QN frequencies for Dirac perturbations, setting $k=1$, $m=2$ , $%
\protect\xi =5$ and varying the set $\{ \Lambda, q, n \}$ taking the ModMax
parameter $\protect\gamma=0.1$.}\centering
\resizebox{0.95\columnwidth}{!}{\begin{tabular}{ccccccccc}
\hline
$n$ & $\Lambda$ & $\omega(q=0.0)$ &  $\omega(q=0.1)$ & $\omega(q=0.2)$ & $\omega(q=0.3)$ & $\omega(q=0.4)$ & $\omega(q=0.5)$ & $\omega(q=0.6)$\\
\hline
0 & 0.00 & 0.960215\, -0.0962564 i & 0.961671\, -0.0962833 i & 0.966106\, -0.0964475 i & 0.973698\, -0.0966992 i & 0.984776\, -0.0969827 i & 0.999889\, -0.0974124 i & 1.01984\, -0.0976272 i 
\\
0 & 0.01 & 0.916158\, -0.0918083 i & 0.917685\, -0.0918681 i & 0.922327\, -0.0920461 i & 0.93027\, -0.0923387 i & 0.941855\, -0.0927369 i & 0.957631\, -0.0932227 i & 0.978453\, -0.0937583 i 
\\
0 & 0.02 & 0.869837\, -0.0871387 i & 0.871445\, -0.0872113 i & 0.876329\, -0.0874283 i & 0.884681\, -0.0877868 i & 0.89685\, -0.0882797 i & 0.913396\, -0.0888918 i & 0.935193\, -0.0895898 i 
\\
0 & 0.03 & 0.820869\, -0.0822089 i & 0.822572\, -0.0822961 i & 0.827743\, -0.0825571 i & 0.836577\, -0.08299 i & 0.849431\, -0.0835893 i & 0.866876\, -0.0843427 i & 0.889806\, -0.0852211 i 
\\
0 & 0.04 & 0.768749\, -0.0769683 i & 0.770566\, -0.0770723 i & 0.776083\, -0.0773841 i & 0.785496\, -0.0779024 i & 0.799169\, -0.0786235 i & 0.817682\, -0.0795373 i & 0.841947\, -0.0806194 i 
\\
0 & 0.05 & 0.712784\, -0.0713475 i & 0.714744\, -0.0714715 i & 0.720687\, -0.0718432 i & 0.730813\, -0.0724621 i & 0.745487\, -0.0733255 i & 0.765298\, -0.0744256 i & 0.791166\, -0.0757415 i 
\\
0 & 0.06 & 0.651986\, -0.0652474 i & 0.654128\, -0.0653958 i & 0.660617\, -0.0658406 i & 0.671649\, -0.0665816 i & 0.687587\, -0.0676165 i & 0.709016\, -0.0689386 i & 0.736862\, -0.0705295 i 
\\
0 & 0.07 & 0.584848\, -0.0585172 i & 0.587235\, -0.0586967 i & 0.594456\, -0.0592349 i & 0.606694\, -0.0601313 i & 0.624294\, -0.0613809 i & 0.647823\, -0.0629773 i & 0.678187\, -0.0649023 i 
\\
0 & 0.08 & 0.508867\, -0.0509062 i & 0.51161\, -0.05113 i & 0.519884\, -0.0517964 i & 0.533838\, -0.052901 i & 0.553763\, -0.0544374 i & 0.580164\, -0.0563887 i & 0.613888\, -0.0587412 i 
\\
0 & 0.09 & 0.419262\, -0.0419362 i & 0.422586\, -0.0422003 i & 0.432573\, -0.043089 i & 0.449249\, -0.0444692 i & 0.472766\, -0.0464741 i & 0.503442\, -0.0489416 i & 0.541969\, -0.0518527 i 
\\
 0 & 0.10 &0.304224\, -0.030426 i & 0.308888\, -0.0317796 i & 0.322427\, -0.0330684 i & 0.344333\, -0.0333962 i & 0.374452\, -0.0355957 i & 0.412665\, -0.0405061 i & 0.458857\, -0.0438653 i 
\\
\hline 
1 & 0.00 & 0.949593\, -0.290179 i & 0.951027\, -0.290182 i & 0.95556\, -0.290741 i & 0.963288\, -0.291546 i & 0.974464\, -0.292271 i & 0.989903\, -0.293737 i & 1.00984\, -0.293718 i 
 \\
1 & 0.01 & 0.906891\, -0.27655 i & 0.908431\, -0.276728 i & 0.913111\, -0.277258 i & 0.921122\, -0.278128 i & 0.932813\, -0.279311 i & 0.948745\, -0.28075 i & 0.96979\, -0.282329 i 
 \\
1 & 0.02 &  0.861871\, -0.262294 i & 0.863485\, -0.262511 i & 0.86839\, -0.263161 i & 0.87678\, -0.264235 i & 0.889011\, -0.265709 i & 0.905653\, -0.267538 i & 0.9276\, -0.269618 i 
\\
1 & 0.03 & 0.814144\, -0.247293 i & 0.815848\, -0.247555 i & 0.821023\, -0.24834 i & 0.829867\, -0.24964 i & 0.842743\, -0.25144 i & 0.860231\, -0.2537 i & 0.88324\, -0.256332 i 
\\
1 & 0.04 & 0.763201\, -0.231394 i & 0.765015\, -0.231707 i & 0.77052\, -0.232645 i & 0.779918\, -0.234204 i & 0.793575\, -0.236373 i & 0.812082\, -0.239121 i & 0.836362\, -0.242372 i 
\\
1 & 0.05 & 0.708342\, -0.214385 i & 0.710294\, -0.214758 i & 0.716213\, -0.215877 i & 0.726301\, -0.21774 i & 0.740928\, -0.220339 i & 0.760688\, -0.223649 i & 0.786515\, -0.227608 i 
\\
 1 & 0.06 & 0.648571\, -0.195969 i & 0.650701\, -0.196415 i & 0.657155\, -0.197754 i & 0.668131\, -0.199984 i & 0.683994\, -0.203098 i & 0.705335\, -0.207076 i & 0.733089\, -0.211863 i 
\\
1 & 0.07 & 0.582371\, -0.175689 i & 0.584744\, -0.176229 i & 0.591921\, -0.177846 i & 0.60409\, -0.180544 i & 0.621593\, -0.184301 i & 0.645003\, -0.189103 i & 0.675233\, -0.194892 i 
\\
1 & 0.08 & 0.507228\, -0.152792 i & 0.509956\, -0.153468 i & 0.518182\, -0.15547 i & 0.532055\, -0.158785 i & 0.551869\, -0.163404 i & 0.578125\, -0.169261 i & 0.61169\, -0.176345 i 
\\
1 & 0.09 & 0.41834\, -0.125841 i & 0.4216\, -0.126543 i & 0.431583\, -0.129296 i & 0.448101\, -0.133293 i & 0.471592\, -0.139487 i & 0.502133\, -0.146931 i & 0.54045\, -0.155628 i 
\\
1 & 0.10 & 0.303869\, -0.0912867 i & 0.310273\, -0.0985442 i & 0.323833\, -0.102539 i & 0.342541\, -0.0976416 i & 0.371697\, -0.102469 i & 0.412673\, -0.122977 i & 0.457876\, -0.131521 i 
\\
\hline 2 & 0.00 & 0.929979\, -0.487634 i & 0.931262\, -0.487429 i & 0.936082\, -0.488541 i & 0.944152\, -0.490025 i & 0.95535\, -0.490895 i & 0.97168\, -0.493824 i & 0.990699\, -0.491978 i 
\\
2 & 0.01 & 0.889651\, -0.464109 i & 0.891213\, -0.464402 i & 0.895963\, -0.465275 i & 0.9041\, -0.466707 i & 0.915986\, -0.468649 i & 0.932201\, -0.471003 i & 0.953654\, -0.473565 i 
\\
2 & 0.02 & 0.846942\, -0.439644 i & 0.848568\, -0.440007 i & 0.853511\, -0.441089 i & 0.861973\, -0.442873 i & 0.87432\, -0.445323 i & 0.891144\, -0.448354 i & 0.913367\, -0.45179 i 
\\
2 & 0.03 & 0.801453\, -0.414041 i & 0.803159\, -0.414479 i & 0.808342\, -0.415792 i & 0.817206\, -0.417966 i & 0.830124\, -0.420973 i & 0.847696\, -0.424746 i & 0.870856\, -0.429132 i 
\\
2 & 0.04 & 0.752658\, -0.387036 i & 0.754465\, -0.387561 i & 0.75995\, -0.389134 i & 0.769321\, -0.391748 i & 0.782953\, -0.395382 i & 0.801452\, -0.399983 i & 0.825766\, -0.405422 i 
\\
2 & 0.05 & 0.699844\, -0.358275 i & 0.701781\, -0.358901 i & 0.707655\, -0.360777 i & 0.717674\, -0.363901 i & 0.732214\, -0.368258 i & 0.751883\, -0.373806 i & 0.777636\, -0.380438 i 
\\
2 & 0.06 & 0.641995\, -0.327251 i & 0.644104\, -0.328 i & 0.650492\, -0.330243 i & 0.661363\, -0.33398 i & 0.677086\, -0.339199 i & 0.698263\, -0.345866 i & 0.725846\, -0.353885 i 
\\
2 & 0.07 & 0.577573\, -0.293201 i & 0.579918\, -0.294104 i & 0.58701\, -0.296808 i & 0.599049\, -0.301328 i & 0.616366\, -0.307614 i & 0.639552\, -0.315654 i & 0.66953\, -0.325348 i 
\\
2 & 0.08 & 0.504033\, -0.25486 i & 0.506738\, -0.255998 i & 0.514869\, -0.259341 i & 0.52858\, -0.264874 i & 0.548184\, -0.272598 i & 0.574151\, -0.282369 i & 0.607428\, -0.294251 i 
\\
2 & 0.09 & 0.416533\, -0.209823 i & 0.419547\, -0.210749 i & 0.429635\, -0.215573 i & 0.445649\, -0.221848 i & 0.46932\, -0.232659 i & 0.499641\, -0.245178 i & 0.537491\, -0.259577 i 
\\
 2 & 0.10 & 0.30317\, -0.152169 i & 0.31721\, -0.172486 i & 0.330974\, -0.179487 i & 0.335725\, -0.155595 i & 0.360794\, -0.158399 i & 0.414552\, -0.208882 i & 0.455814\, -0.218998 i 
 \\
\hline  
3 & 0.00 & 0.903578\, -0.689241 i & 0.904441\, -0.688644 i & 0.90986\, -0.69047 i & 0.918574\, -0.692761 i & 0.9295\, -0.693482 i & 0.94761\, -0.698322 i & 0.96362\, -0.692986 i 
\\
3 & 0.01 & 0.866171\, -0.654985 i & 0.867764\, -0.655392 i & 0.872612\, -0.656603 i & 0.880921\, -0.658585 i & 0.893073\, -0.661267 i & 0.909676\, -0.664504 i & 0.931681\, -0.668 i 
\\
3 & 0.02 & 0.826383\, -0.619585 i & 0.828027\, -0.620092 i & 0.833026\, -0.621608 i & 0.841591\, -0.624107 i & 0.854105\, -0.627532 i & 0.871185\, -0.631761 i & 0.893794\, -0.636537 i 
\\
3 & 0.03 & 0.78379\, -0.582754 i & 0.7855\, -0.583371 i & 0.790699\, -0.585219 i & 0.799599\, -0.588279 i & 0.812587\, -0.592509 i & 0.830285\, -0.597811 i & 0.853667\, -0.603961 i 
\\
3 & 0.04 & 0.737839\, -0.544118 i & 0.739638\, -0.544858 i & 0.745101\, -0.547077 i & 0.754442\, -0.550764 i & 0.76805\, -0.555888 i & 0.786552\, -0.562371 i & 0.810929\, -0.570029 i 
\\
3 & 0.05 &  0.687788\, -0.503171 i & 0.689704\, -0.504054 i & 0.695521\, -0.506701 i & 0.70545\, -0.511108 i & 0.719878\, -0.517253 i & 0.739433\, -0.525078 i & 0.765098\, -0.534427 i 
\\
3 & 0.06 & 0.632584\, -0.459197 i & 0.634662\, -0.460251 i & 0.64096\, -0.463412 i & 0.651688\, -0.468679 i & 0.667222\, -0.476036 i & 0.688179\, -0.485433 i & 0.715534\, -0.496734 i 
\\
3 & 0.07 & 0.570648\, -0.411114 i & 0.572953\, -0.412382 i & 0.579924\, -0.416182 i & 0.591785\, -0.422551 i & 0.608841\, -0.431394 i & 0.631717\, -0.442714 i & 0.661347\, -0.456362 i 
\\
3 & 0.08 & 0.499385\, -0.35714 i & 0.502066\, -0.358752 i & 0.510059\, -0.363443 i & 0.52353\, -0.371202 i & 0.542836\, -0.38206 i & 0.568375\, -0.39576 i & 0.601277\, -0.412518 i 
\\
3 & 0.09 & 0.413886\, -0.293896 i & 0.416327\, -0.294817 i & 0.426769\, -0.301934 i & 0.441703\, -0.310121 i & 0.466046\, -0.326013 i & 0.496128\, -0.343719 i & 0.53319\, -0.363733 i 
\\
3 & 0.10 & 0.30214\, -0.213074 i & 0.333779\, -0.253119 i & 0.348112\, -0.263422 i & 0.31933\, -0.205957 i & 0.333515\, -0.200038 i & 0.42048\, -0.298349 i & 0.452546\, -0.306284 i 
\\
\hline
\end{tabular}
\label{table:Fourth_set}
}
\end{table*}

\begin{table*}[tbp]
\caption{QN frequencies for Dirac perturbations, setting $k=1$, $m=2$ , $%
\protect\xi =5$ and varying the set $\{ \Lambda, q, n \}$ taking the ModMax
parameter $\protect\gamma=0.2$. }\centering
\resizebox{0.95\columnwidth}{!}{\begin{tabular}{ccccccccc}
\hline
$n$ & $\Lambda$ & $\omega(q=0.0)$ &  $\omega(q=0.1)$ & $\omega(q=0.2)$ & $\omega(q=0.3)$ & $\omega(q=0.4)$ & $\omega(q=0.5)$ & $\omega(q=0.6)$\\
\hline
0 & 0.00 &  0.960215\, -0.0962564 i & 0.961534\, -0.0963004 i & 0.965538\, -0.0964247 i & 0.972377\, -0.0966365 i & 0.982315\, -0.0968823 i & 0.995791\, -0.0972709 i & 1.01344\, -0.0976058 i \\
0 & 0.01 & 0.916158\, -0.0918083 i & 0.917539\, -0.0918624 i & 0.921733\, -0.0920236 i & 0.92889\, -0.092289 i & 0.939287\, -0.0926516 i & 0.953358\, -0.0930978 i & 0.97177\, -0.0936 i \\
0 & 0.02 & 0.869837\, -0.0871387 i & 0.871291\, -0.0872044 i & 0.875704\, -0.0874008 i & 0.883231\, -0.0877257 i & 0.894153\, -0.0881735 i & 0.908917\, -0.0887329 i & 0.928202\, -0.0893798 i \\
0 & 0.03 & 0.820869\, -0.0822089 i & 0.822409\, -0.0822878 i & 0.827082\, -0.0825241 i & 0.835044\, -0.0829161 i & 0.846585\, -0.0834598 i & 0.862157\, -0.0841459 i & 0.882458\, -0.084954 i \\
0 & 0.04 & 0.768749\, -0.0769683 i & 0.770393\, -0.0770624 i & 0.775377\, -0.0773446 i & 0.783864\, -0.0778138 i & 0.796143\, -0.0784673 i & 0.812679\, -0.0792977 i & 0.834179\, -0.0802878 i \\
0 & 0.05 & 0.712784\, -0.0713475 i & 0.714557\, -0.0714597 i & 0.719928\, -0.071796 i & 0.729058\, -0.0723562 i & 0.742244\, -0.0731382 i & 0.759951\, -0.0741363 i & 0.782897\, -0.0753362 i \\
0 & 0.06 & 0.651986\, -0.0652474 i & 0.653923\, -0.0653816 i & 0.659788\, -0.0657843 i & 0.669739\, -0.0664548 i & 0.684069\, -0.0673917 i & 0.703242\, -0.0685904 i & 0.727976\, -0.070038 i \\
0 & 0.07 & 0.584848\, -0.0585172 i & 0.587008\, -0.0586798 i & 0.593535\, -0.0591672 i & 0.604579\, -0.0599775 i & 0.620417\, -0.0611096 i & 0.641497\, -0.0625569 i & 0.668521\, -0.064307 i \\
0 & 0.08 & 0.508867\, -0.0509062 i & 0.511349\, -0.051109 i & 0.51883\, -0.0517083 i & 0.531433\, -0.0527154 i & 0.549387\, -0.0541002 i & 0.573091\, -0.0558777 i & 0.603192\, -0.0580113 i \\
0 & 0.09 & 0.419262\, -0.0419362 i & 0.422279\, -0.042273 i & 0.431308\, -0.04301 i & 0.446393\, -0.0442888 i & 0.467633\, -0.0460763 i & 0.495268\, -0.0482607 i & 0.529821\, -0.0509556 i \\
0 & 0.10 & 0.304224\, -0.030426 i & 0.308966\, -0.0363733 i & 0.320123\, -0.0264265 i & 0.340761\, -0.0347901 i & 0.368067\, -0.0362689 i & 0.402569\, -0.038795 i & 0.44442\, -0.0425248 i 
\\
\hline 
1 & 0.00 & 0.949593\, -0.290179 i & 0.95093\, -0.29031 i & 0.954976\, -0.290656 i & 0.961906\, -0.291278 i & 0.971905\, -0.291854 i & 0.985657\, -0.293152 i & 1.0035\, -0.293991 i \\
1 & 0.01 & 0.906891\, -0.27655 i & 0.908284\, -0.276711 i & 0.912512\, -0.277191 i & 0.91973\, -0.27798 i & 0.930221\, -0.279057 i & 0.944428\, -0.28038 i & 0.963034\, -0.281863 i \\
1 & 0.02 & 0.861871\, -0.262294 i & 0.863331\, -0.262491 i & 0.867762\, -0.263079 i & 0.875323\, -0.264052 i & 0.8863\, -0.265392 i & 0.901147\, -0.267063 i & 0.920559\, -0.268993 i \\
1 & 0.03 & 0.814144\, -0.247293 i & 0.815686\, -0.24753 i & 0.820361\, -0.24824 i & 0.828332\, -0.249418 i & 0.839891\, -0.251051 i & 0.855499\, -0.25311 i & 0.875864\, -0.255532 i \\
1 & 0.04 & 0.763201\, -0.231394 i & 0.764842\, -0.231677 i & 0.769816\, -0.232526 i & 0.778287\, -0.233938 i & 0.790552\, -0.235903 i & 0.807079\, -0.238401 i & 0.828587\, -0.241376 i \\
1 & 0.05 & 0.708342\, -0.214385 i & 0.710107\, -0.214723 i & 0.715456\, -0.215735 i & 0.724552\, -0.217421 i & 0.737694\, -0.219775 i & 0.755353\, -0.222779 i & 0.778255\, -0.226389 i \\
1 & 0.06 & 0.648571\, -0.195969 i & 0.650498\, -0.196372 i & 0.656331\, -0.197584 i & 0.666231\, -0.199602 i & 0.680492\, -0.202421 i & 0.699583\, -0.206029 i & 0.724229\, -0.210384 i \\
1 & 0.07 & 0.582371\, -0.175689 i & 0.584518\, -0.176178 i & 0.591007\, -0.177645 i & 0.601986\, -0.18008 i & 0.617736\, -0.183485 i & 0.638708\, -0.187838 i & 0.665608\, -0.193103 i \\
1 & 0.08 & 0.507228\, -0.152792 i & 0.509697\, -0.153406 i & 0.517127\, -0.155192 i & 0.529669\, -0.158239 i & 0.547509\, -0.162375 i & 0.571094\, -0.167735 i & 0.601038\, -0.174138 i \\
1 & 0.09 & 0.41834\, -0.125841 i & 0.421481\, -0.127111 i & 0.43038\, -0.129164 i & 0.445372\, -0.132967 i & 0.466529\, -0.138378 i & 0.493932\, -0.144756 i & 0.528367\, -0.152948 i \\
1 & 0.10 & 0.303869\, -0.0912867 i & 0.320049\, -0.126953 i & 0.311094\, -0.0585326 i & 0.342185\, -0.107868 i & 0.367555\, -0.10893 i & 0.401061\, -0.114813 i & 0.443177\, -0.12686 i 
\\
\hline 
2 & 0.00 & 0.929979\, -0.487634 i & 0.931374\, -0.487896 i & 0.935443\, -0.488358 i & 0.942541\, -0.489371 i & 0.952352\, -0.489721 i & 0.966921\, -0.492412 i & 0.984936\, -0.493353 i \\
2 & 0.01 & 0.889651\, -0.464109 i & 0.891064\, -0.464374 i & 0.895355\, -0.465165 i & 0.902686\, -0.466464 i & 0.913349\, -0.468234 i & 0.927806\, -0.470399 i & 0.946763\, -0.472813 i \\
2 & 0.02 & 0.846942\, -0.439644 i & 0.848413\, -0.439972 i & 0.852879\, -0.440952 i & 0.860503\, -0.44257 i & 0.871582\, -0.444796 i & 0.886586\, -0.447569 i & 0.906232\, -0.45076 i \\
2 & 0.03 & 0.801453\, -0.414041 i & 0.802996\, -0.414438 i & 0.807679\, -0.415625 i & 0.815667\, -0.417595 i & 0.827262\, -0.420324 i & 0.842938\, -0.423762 i & 0.863426\, -0.4278 i \\
2 & 0.04 & 0.752658\, -0.387036 i & 0.754292\, -0.387512 i & 0.759249\, -0.388935 i & 0.767695\, -0.391301 i & 0.779935\, -0.394595 i & 0.796449\, -0.398777 i & 0.817975\, -0.403756 i \\
2 & 0.05 & 0.699844\, -0.358275 i & 0.701596\, -0.358841 i & 0.706904\, -0.360539 i & 0.715937\, -0.363366 i & 0.728998\, -0.367313 i & 0.74657\, -0.372347 i & 0.769395\, -0.378396 i \\
2 & 0.06 & 0.641995\, -0.327251 i & 0.643902\, -0.327927 i & 0.649676\, -0.329959 i & 0.65948\, -0.333339 i & 0.673612\, -0.338064 i & 0.692552\, -0.344109 i & 0.717035\, -0.351406 i \\
2 & 0.07 & 0.577573\, -0.293201 i & 0.579696\, -0.294022 i & 0.586111\, -0.296476 i & 0.596964\, -0.300547 i & 0.612553\, -0.306256 i & 0.633313\, -0.313533 i & 0.65998\, -0.322348 i \\
2 & 0.08 & 0.504033\, -0.25486 i & 0.506488\, -0.255905 i & 0.513786\, -0.258823 i & 0.526248\, -0.263991 i & 0.543823\, -0.270815 i & 0.567227\, -0.279871 i & 0.596836\, -0.290508 i \\
2 & 0.09 & 0.416533\, -0.209823 i & 0.419721\, -0.211577 i & 0.428448\, -0.215148 i & 0.443237\, -0.221455 i & 0.464315\, -0.230654 i & 0.491132\, -0.241016 i & 0.525597\, -0.255232 i \\
2 & 0.10 & 0.30317\, -0.152169 i & 0.316763\, -0.172249 i & 0.285112\, +0.0692093 i & 0.349571\, -0.188813 i & 0.366687\, -0.181886 i & 0.394612\, -0.183928 i & 0.438893\, -0.207583 i 
 \\
\hline 
3 & 0.00 & 0.903578\, -0.689241 i & 0.905072\, -0.689637 i & 0.9091\, -0.69015 i & 0.916423\, -0.691528 i & 0.925507\, -0.691249 i & 0.941686\, -0.695683 i & 0.959551\, -0.696324 i \\
3 & 0.01 & 0.866171\, -0.654985 i & 0.867612\, -0.655353 i & 0.871991\, -0.65645 i & 0.879477\, -0.658249 i & 0.890376\, -0.660694 i & 0.905173\, -0.663676 i & 0.924608\, -0.666978 i \\
3 & 0.02 & 0.826383\, -0.619585 i & 0.82787\, -0.620044 i & 0.832386\, -0.621416 i & 0.840102\, -0.623681 i & 0.851328\, -0.626795 i & 0.866555\, -0.630666 i & 0.886531\, -0.635108 i \\
3 & 0.03 & 0.78379\, -0.582754 i & 0.785337\, -0.583312 i & 0.790034\, -0.584985 i & 0.798053\, -0.587757 i & 0.809707\, -0.591596 i & 0.82549\, -0.596428 i & 0.846161\, -0.602095 i \\
3 & 0.04 & 0.737839\, -0.544118 i & 0.739467\, -0.544788 i & 0.744402\, -0.546796 i & 0.752821\, -0.550133 i & 0.765035\, -0.554778 i & 0.781544\, -0.560673 i & 0.80311\, -0.567685 i \\
3 & 0.05 & 0.687788\, -0.503171 i & 0.689521\, -0.50397 i & 0.694776\, -0.506364 i & 0.703727\, -0.510353 i & 0.716685\, -0.515921 i & 0.734146\, -0.523021 i & 0.756877\, -0.53155 i \\
3 & 0.06 & 0.632584\, -0.459197 i & 0.634462\, -0.460148 i & 0.640157\, -0.463013 i & 0.649828\, -0.467776 i & 0.663786\, -0.474435 i & 0.682522\, -0.482957 i & 0.706786\, -0.493239 i \\
3 & 0.07 & 0.570648\, -0.411114 i & 0.572739\, -0.41227 i & 0.579052\, -0.415725 i & 0.589722\, -0.421445 i & 0.605091\, -0.429489 i & 0.625554\, -0.439725 i & 0.651899\, -0.452137 i \\
3 & 0.08 & 0.499385\, -0.35714 i & 0.501834\, -0.358631 i & 0.508899\, -0.36265 i & 0.521301\, -0.37001 i & 0.538431\, -0.379475 i & 0.561635\, -0.39231 i & 0.590732\, -0.40719 i \\
3 & 0.09 & 0.413886\, -0.293896 i & 0.417469\, -0.296666 i & 0.425721\, -0.301426 i & 0.440146\, -0.31016 i & 0.461222\, -0.323279 i & 0.486813\, -0.337201 i & 0.521631\, -0.357762 i \\
3 & 0.10 & 0.30214\, -0.213074 i & 0.333302\, -0.252763 i & 0.368464\, +0.290771 i & 0.367417\, -0.277134 i & 0.365654\, -0.255162 i & 0.38039\, -0.247944 i & 0.430334\, -0.286034 i 
\\
\hline
\end{tabular}
\label{table:Fifth_set}
}
\end{table*}

\section{Quasinormal modes and Lyapunov exponent}

In this part, we are interested in exploring the QNMs in large $\ell$ limit
called eikonal limits $(\ell \longrightarrow\infty)$. In Ref. \cite{cardoso}, it
has been discussed that in a static and spherically symmetric spacetime,
both real and imaginary parts of QNMs are related to the characteristics of
circular null geodesics, via the angular velocity and the Lyapunov exponent
of the photon sphere correspond, respectively, as follows 
\begin{equation}  \label{qnm}
\omega(\ell \gg 1) = \Omega_c  \ell -i \bigg(n+\frac{1}{2}\bigg)|\lambda_L|,
\end{equation}
here, $\Omega_c$ and $\lambda_L$ are the angular velocity at the unstable
null geodesic, and the Lyapunov exponent, respectively. $n$ indicates the
overtone. To obtain the radius of unstable circular orbit or photonic orbit $%
r_{ph}$, first, we investigate the null geodesics via Hamiltonian of a free
photon moving in the black hole background 
\begin{equation}
H=\frac{1}{2}g^{ij}p_i p_j=0.
\end{equation}

Without the loss of generosity, the trajectory in the equatorial plane will
be considered where $\theta =\frac{\pi }{2}$ which leads to the following
equation 
\begin{equation}
-\frac{p_{t}^{2}}{f(r)}+f(r)p_{r}^{2}+\frac{p_{\phi }^{2}}{r^{2}}=0.
\end{equation}

As the Hamiltonian does not depend on the $t$ and $\phi $ coordinates, we
define two constants of motion, $E=-p_{t}$ and $L=p_{\phi }$ as energy and
angular momentum respectively. Applying the Hamiltonian formalism 
\begin{equation}
\dot{r}=\frac{\partial H}{\partial p_{r}}=p_{r}f(r).
\end{equation}

Using the definition of the effective potential $\dot{r}^{2}=V_{eff}(r)=0$
and the two conserved quantities $\{E,L\}$, then we can write the effective
potential according to 
\begin{equation}
V_{eff}(r)=E^{2}-f(r)\frac{L^{2}}{r^{2}}\,.
\end{equation}

The conditions $V_{eff}(r)=0$ and $V_{eff}^{\prime }(r)=0$ for circular null
geodesics lead, respectively, to: 
\begin{equation}
\frac{E}{L}=\pm \frac{1}{r_{1}}\sqrt{f(r_{1})}\,,  \label{LElight}
\end{equation}%
and 
\begin{equation}
2f(r_{1})-r_{1}\frac{df(r)}{dr}\Bigg|_{r_{1}}=0,  \label{cirgeo}
\end{equation}%
the last equation, Eq. \eqref{cirgeo}, is precisely required to obtain the
critical value $r_{1}$ or $r_{ph}$.

Thus, with the help of \eqref{cirgeo}, we obtain $r_{ph}$ as follows 
\begin{equation}
r_{ph}=\frac{3m}{4}+\frac{\sqrt{9m^{2}-32q^{2}e^{-\gamma }}}{4}.
\end{equation}

Now, remembering that $\lambda _{L}$ and $\Omega _{c}$ are, respectively,
the Lyapunov exponent and the coordinate angular velocity at the unstable
null geodesic (defined according to \cite{cardoso}), we can compute directly
by means of the expressions 
\begin{align}
\lambda _{L}& \equiv \sqrt{\frac{f(r_{1})r_{1}^{2}}{2}\Bigg(\frac{\mathrm{%
d^{2}}}{\mathrm{d}r^{2}}\frac{f(r)}{r^{2}}\Bigg)\Bigg|_{r=r_{1}}}=\pm
r_{1}^{2}\sqrt{\frac{g^{\prime \prime }(r_{1})g(r_{1})}{2}},  \label{Lambda}
\\
&  \notag \\
\Omega _{c}& \equiv \frac{\dot{\phi}(r_{1})}{\dot{t}(r_{1})}=\frac{\sqrt{%
f(r_{1})}}{r_{1}}=\sqrt{g(r_{1})}.  \label{Omega}
\end{align}%
Also, should be mentioned that $\lambda _{L}$ parametrize the rate of
convergence or divergence of null rays in the ring's vicinity, i.e., $%
\lambda _{L}$ is the decay rate of the unstable circular null geodesics. The
(square of the) first parameter is then given according to 
\begin{align}
\lambda _{L}^{2}& =\frac{\sqrt{9m^{2}-32q^{2}e^{-\gamma }}}{%
16q^{2}e^{-\gamma }}\left( \frac{9m^{3}}{32q^{6}e^{-3\gamma }}\left( \frac{%
9m^{2}}{8}+7q^{2}e^{-\gamma }\right) +\frac{m}{q^{2}e^{-\gamma }}\left( 
\frac{11}{4}+\Lambda q^{2}e^{-\gamma }\right) \right)   \notag \\
&  \notag \\
& +\frac{243m^{4}}{512q^{8}e^{-4\gamma }}\left( q^{2}e^{-\gamma }-\frac{m^{2}%
}{8}\right) -\frac{3\Lambda m^{2}}{16q^{4}e^{-2\gamma }}\left( \frac{23}{4}%
+\Lambda q^{2}e^{-\gamma }\right) +\frac{8}{16q^{2}e^{-\gamma }}-\frac{%
2\Lambda }{3},
\end{align}%
and the second parameter is computed to be 
\begin{equation}
\Omega _{c}=\frac{\sqrt{\frac{m\sqrt{9m^{2}-32q^{2}e^{-\gamma }}}{%
2q^{2}e^{-\gamma }}\left( 1-4\Lambda q^{2}e^{-\gamma }\right) -\frac{3m^{2}}{%
2q^{2}e^{-\gamma }}\left( 1+4\Lambda q^{2}e^{-\gamma }\right) +\frac{%
32\Lambda q^{2}e^{-\gamma }}{3}+8}}{\sqrt{9m^{2}-32q^{2}e^{-\gamma }}+3m}.
\end{equation}

Finally, let us mention that the last two expressions recover the standard
results when we demand $\gamma \rightarrow 0$, and also contains the neutral
case, recovered when $q \rightarrow 0$, as it should be.

\section{Emission Rate}

Quantum fluctuations inside the black holes, create and annihilate a large
number of particles near the horizon. Particles with positive energy can
escape from black hole via tunneling effect. This phenomenon causes gradual
black hole evaporation over a period of time known as Hawking radiation. For
a distant observer, black hole shadow corresponds to the high energy
absorption cross section which is shown that it is around a limiting
constant value \cite{wei} 
\begin{equation}
\sigma_{lim}\approx \pi R_{sh}^2
\end{equation}
where $R_{sh}$ is the shadow radius. Then, the expression for the rate of
energy emission can be written as \cite{wei,pap,EslamPanah,hend,ali} 
\begin{equation}  \label{emission}
\frac{{{d^2}E}}{{d\omega dt}} = \frac{{2{\pi ^2}\sigma_{lim} }}{{{e^{\frac{%
\omega }{T_H}}} - 1}}{\omega ^3}
\end{equation}

To obtain the shadow radius, The radius of shadow detected by an observer at
spatial infinity is related to the critical orbit and the observer's
position $r_{O}$ as \cite{volker} 
\begin{align}
{R_{sh}}& ={r_{c}}\sqrt{\frac{{f({r_{O}})}}{{f({r_{c}})}}}  \notag \\
&  \notag \\
& =\frac{\left( \sqrt{9m^{2}-32q^{2}e^{-\gamma }}+3m\right) \sqrt{1-\frac{m}{%
r_{O}}+\frac{q^{2}e^{-\gamma }}{r_{O}^{2}}-\frac{\Lambda r_{O}^{2}}{3}}}{%
\sqrt{\frac{m\sqrt{9m^{2}-32q^{2}e^{-\gamma }}}{2q^{2}e^{-\gamma }}\left(
1-4\Lambda q^{2}e^{-\gamma }\right) -\frac{3m^{2}}{2q^{2}e^{-\gamma }}\left(
1+4\Lambda q^{2}e^{-\gamma }\right) +\frac{8}{3}\left( 3+4\Lambda
q^{2}e^{-\gamma }\right) }}.  \label{Rsh}
\end{align}

By substituting the Hawking temperature from Eq. \eqref{TemII} and $R_{sh}$
from Eq. \eqref{Rsh} in Eq. \eqref{emission}, the energy emission rate can
be calculated. The behavior of energy emission rate with respect to $\omega$
for fixed values of cosmological constant and different values of ModMax
parameter for $m=1$, $q=0.3$ and $r_O=3$ are represented in Fig. \ref%
{fig:Em1}

%%%%%%%%%%%%%%%%%%%%%%%%%%%%%%%%%%%%%%%%%%%%%%%%%%%%%%%%%%%%%%%%%%%
\begin{figure}[tbph]
\centering
\includegraphics[width=0.43\linewidth]{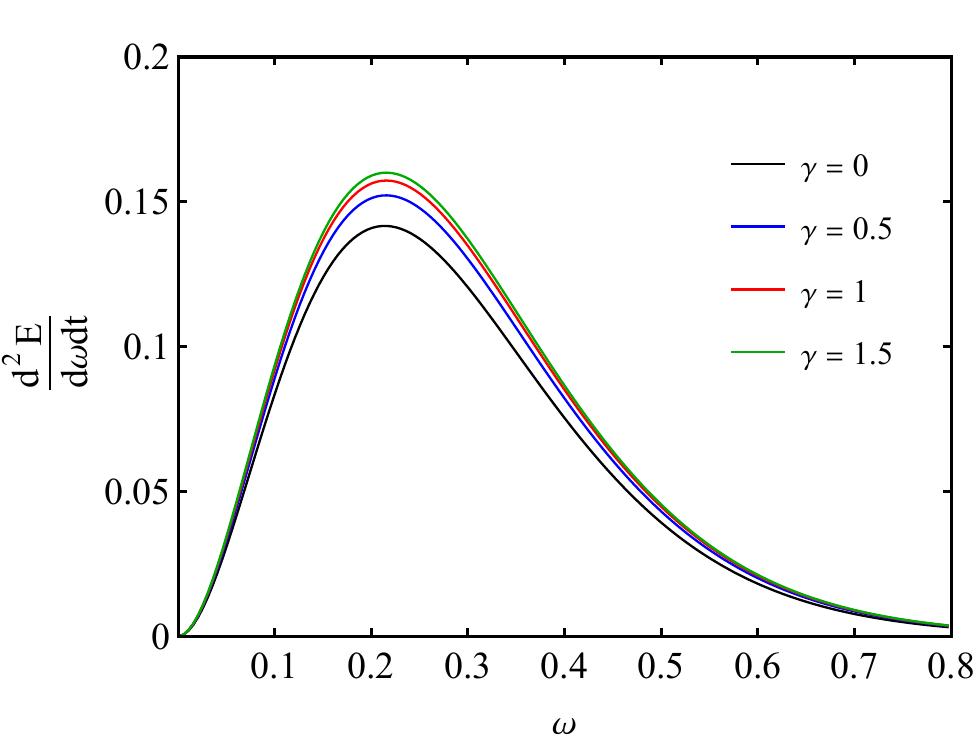} %
\includegraphics[width=0.43\linewidth]{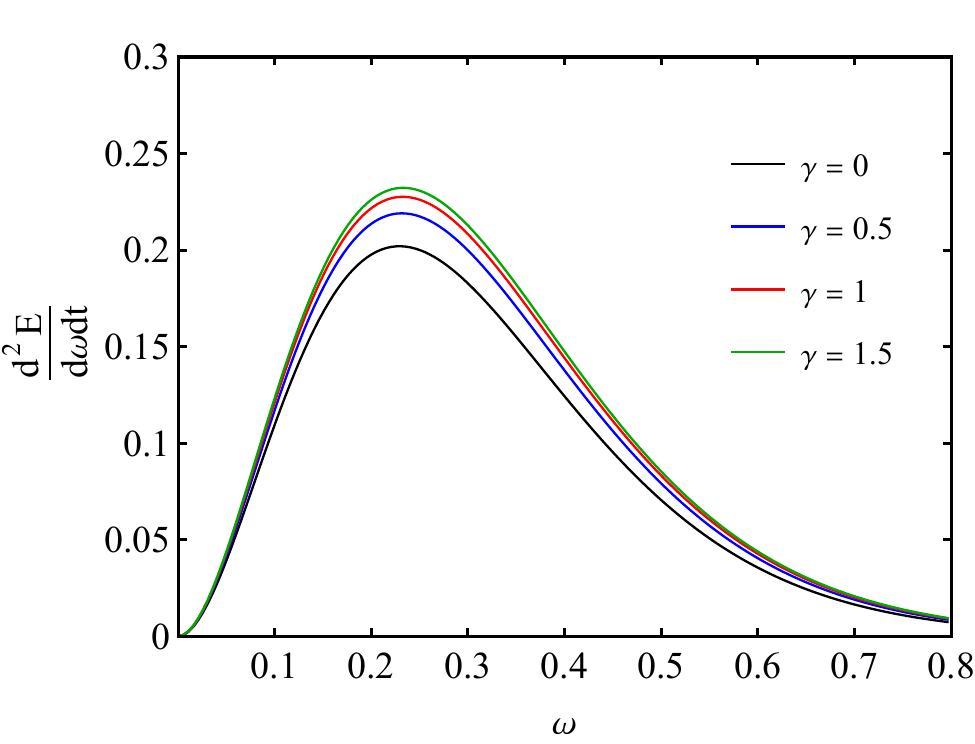} \newline
\caption{The emission rate versus frequency for $k=1$, $m=1$, $q=0.3$ and
different values of the $\protect\gamma$. Left panel for dS case ($%
\Lambda=+0.003$). Right panel for AdS case ($\Lambda=-0.003$).}
\label{fig:Em1}
\end{figure}
%%%%%%%%%%%%%%%%%%%%%%%%%%%%%%%%%%%%%%%%%%%%%%%%%%%%%%%%%%%%%%%%%%

In Fig. \ref{fig:Em1}, left panel shows the dS case with positive
cosmological constant $\Lambda = 0.003$. It can be noticed that when the
ModMax parameter becomes higher the peak of energy emission rate goes up and
the maximum energy emission rate occurs at bigger frequency. The right
panel, demonstrates the AdS case for $\Lambda = - 0.003$. The variation of
emission rate versus frequency for different values of $\gamma$ has the same
behavior as dS case. The bigger values of $\gamma$ leads to higher emission
rate and shift the maximum of plots to higher frequency. This indicates that
the higher ModMax parameter corresponds to the faster evaporation process in
black hole in both dS and AdS spacetime.

The energy emission rate against the frequency are shown for $m=1$, $q=0.3$, 
$\gamma=1$ and different values of the cosmological constant in Fig. \ref%
{fig:Em2}. The left panel shows that the rate of emission energy is become
smaller for higher values of $\Lambda$ in dS spacetime and it means that
increasing the cosmological constant in this spacetime yields to slower
evaporation process and a longer lifetime for the black hole. Moreover, the
peak of the emission rate decreases with increasing $\Lambda$ and shifts to
the lower frequency. In the right panel, the emission rate versus frequency
is described for AdS spacetime and various values of $\Lambda$. When the
absolute value of $\Lambda$ increases the peak of emission rate increases as
well and it shifts to higher right, which means that the maximum emission
rate happens in higher frequency. These results imply that the the
evaporation of the black hole is high for smaller absolute values of $%
\Lambda $ in AdS spacetime. 
%%%%%%%%%%%%%%%%%%%%%%%%%%%%%%%%%%%%%%%%%%%%%%%%%%%%%%%%%%%%%%%%
\begin{figure}[h!]
\centering
\includegraphics[width=0.43\linewidth]{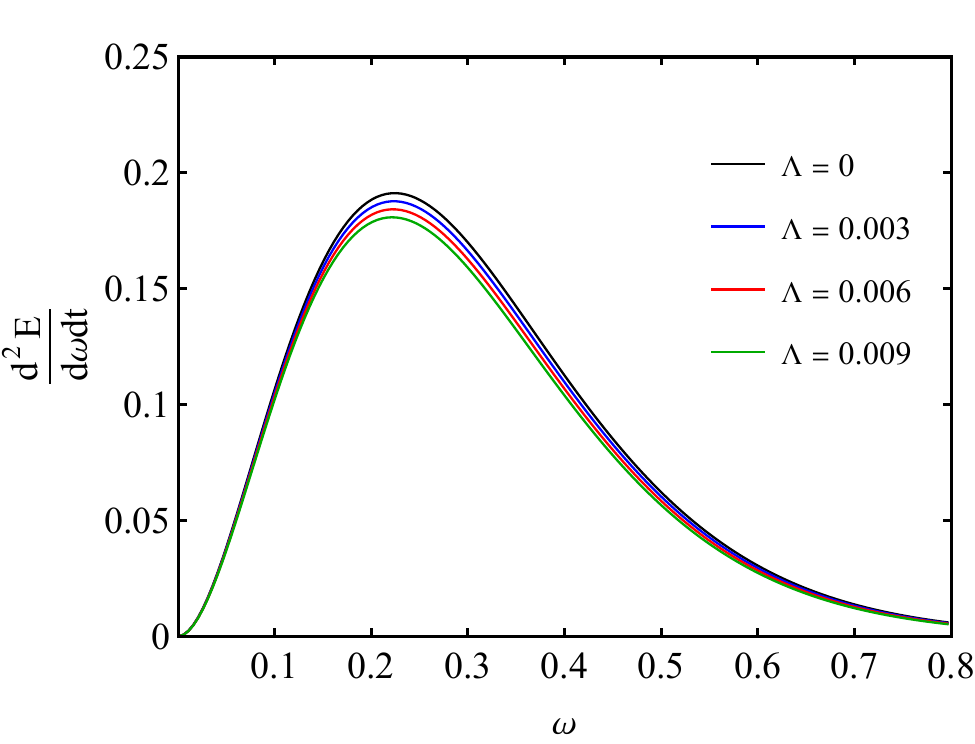} %
\includegraphics[width=0.43\linewidth]{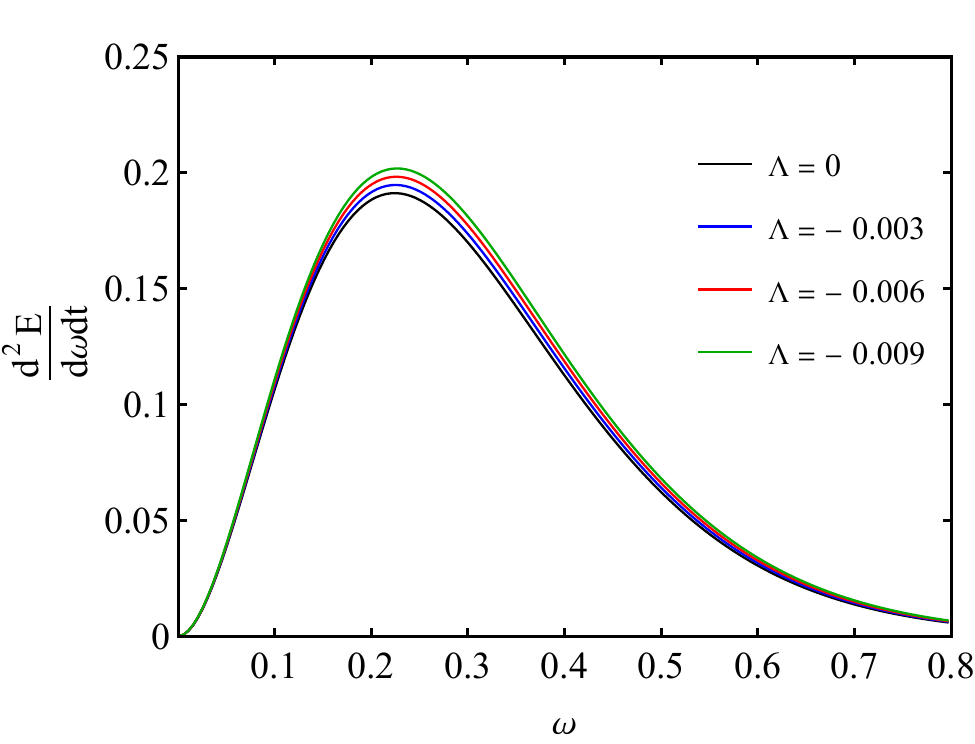} \newline
\caption{The emission rate versus frequency for $k=1$, $m=1$, $q=0.3$, $%
\protect\gamma = 1$ and different values of the $\Lambda$. Left panel for dS
case ($\Lambda>0$). Right panel for AdS case ($\Lambda<0$).}
\label{fig:Em2}
\end{figure}
%%%%%%%%%%%%%%%%%%%%%%%%%%%%%%%%%%%%%%%%%%%%%%%%%%%%%%%%%%%%%%%%

\section{\textbf{Conclusions}}

To summarize our work, in the present paper we have studied several aspects
of topological (A)dS black holes combined with ModMax non-linear
electrodynamics. In particular, we have investigated the quasinormal spectra
for massless scalar, electromagnetic as well as Dirac perturbations in
four-dimensional non-linearly charged dS black holes. We start by revisiting
the basic idea and relevance of the QNMs. Thus, we discuss the method used
to compute the quasinormal frequencies we present the effective potential
and some figures. The QN frequencies were obtained taking advantage WKB
method of 6th order. We present our results in tables, where we show the
impact on the spectrum of the overtone number $n$, the angular degree $\ell$
(or equivalently the parameter $\xi$) and the ModMax parameter $\gamma$. At
least for the numerical values used, all the modes are found to be stable
against the three types of perturbations. Thus, from tables, we can read
that:

\begin{itemize}
\item When the ModMax parameter $\gamma$ goes from zero to positive values,
the quasinormal modes decrease, in other words, fixing $\{ n, \ell ( \text{%
or }\xi) \}$, we notice that as $\gamma$ increases, the QN frequencies also
decrease.

\item Fixing the ModMax parameter $\gamma$, we observe that as $n$
increases, the QN frequencies decrease, regardless of the parameter $\ell ( 
\text{or } \xi)$.

\item In all the cases, the ModMax parameter $\gamma$ does not modify the
stability of the black solution against scalar/electromagnetic/Dirac
perturbations, as is revealed by the minus sign of the imaginary part of the
QN frequencies. Thus, according to this criterion the solution is stable.
\end{itemize}

Moreover, circular null geodesics were studied to investigate the QNMs in
the eikonal limit. By obtaining the angular velocity as the real part and
the Lyapunov exponent as the imaginary component of QNMs, an analytical
expression for QNMs has been derived. Finally, the impact of the
cosmological constant and ModMax parameter on the energy emission rate has
been investigated. It was shown that the rate of energy emission is higher
for the higher value of $\gamma$ for both dS and AdS spacetime, which means
that a higher ModMax parameter makes the black hole's lifetime shorter. For
a fixed value of the ModMax parameter, in dS spacetime, when the
cosmological constant increased the peak of the emission rate decreased, and
for AdS spacetime, high absolute values of $\Lambda $ led to a higher peak
in the emission rate.

%%%%%%%%%%%%%%%%%%%%%%%%%%%%%%%%%

\begin{acknowledgements}
BEP and NH would like to thank University of Mazandaran. 
A. R. acknowledges financial support from Conselleria d'Educació, Cultura, Universitats i Ocupació de la Generalitat Valenciana thorugh Prometeo Project CIPROM/2022/13.
A. R. is funded by the María Zambrano contract ZAMBRANO 21-25 (Spain) (with funding from NextGenerationEU).

\end{acknowledgements}

\end{document}